# Structural Stress as a Predictor of the Rate and Spatial Location of Aortic Growth in Uncomplicated Type B Aortic Dissection

Yuhang Du[1], Yuxuan Wu[2], Hannah L. Cebull[3], Bangquan Liao[1], Rishika Agarwal[4], Alan Meraz[1], Hai Dong[5], Asanish Kalyanasundaram[6], John N. Oshinski[3,4], Rudolph L. Gleason Jr[4,7], John A. Elefteriades[6], Bradley G. Leshnower[5], Minliang Liu[1]

[1] Department of Mechanical Engineering, Texas Tech University, Lubbock, TX

[2] Mercer University School of Medicine, Macon, GA

[3]Department of Radiology & Imaging Science, Emory University, Atlanta, GA

[4] The Wallace H. Coulter Department of Biomedical Engineering

Georgia Institute of Technology and Emory University, Atlanta, GA

[5]Division of Cardiothoracic Surgery, Department of Surgery, Emory University School of Medicine, Atlanta, GA

[6]Aortic Institute at Yale-New Haven Hospital, Yale University School of Medicine, New Haven, CT

[7] The George W. Woodruff School of Mechanical Engineering

Georgia Institute of Technology, Atlanta, GA

Corresponding Authors:

Bradley G. Leshnower, MD

Director of Thoracic Aortic Surgery, Professor of Surgery

Division of Cardiothoracic Surgery, Emory University School of Medicine

1365 Clifton Road, Suite A 2213, Atlanta, GA 30322

Phone: 404.778.3154

Email: bleshno@emory.edu

Minliang Liu, Ph.D.

Assistant Professor, Department of Mechanical Engineering, Texas Tech University

805 Boston Ave, Lubbock, TX 79415

Phone: 806.834.8335

Email: minliang.liu@ttu.edu

## ABSTRACT

Accurate prediction of aortic expansion in uncomplicated type B aortic dissection (TBAD) can help identify patients who may benefit from timely thoracic endovascular aortic repair. To investigate the associations between biomechanical predictors and aortic growth outcomes, the earliest baseline and follow-up CT images from 30 patients with uncomplicated TBAD were obtained in this study. For each patient, a reduced-order fluid-structure interaction (FSI) analysis using the forward penalty stress computation method was performed on the baseline geometry. Aortic growth was quantified by registering baseline and follow-up surfaces using nonrigid registration. Mixed-effects linear and logistic regression analyses were performed to assess the relationships between biomechanical predictors (structural stress, wall shear stress (WSS), and pressure) and aortic growth outcomes while accounting for inter-patient variability. Patients were divided into groups based upon: 1) stable vs enlarging aortic diameters and; 2) treatment with either OMT or surgical intervention, and spatial distributions of biomechanical variables along the dissected aorta were compared. Aortic diameter of the baseline TBAD geometry was also included for comparison. Results of the linear regression revealed a positive association between structural stress and aortic growth rate ($p = 0.0003$) and a negative association for WSS ($p = 0.0227$). Logistic regression yielded area under the receiver operator characteristic curve (AUCs) of 0.7414, 0.5953, 0.4991, and 0.6845 for structural stress, WSS, pressure, and aortic diameter, respectively. Group comparisons showed significant regional differences in structural stress ($p<0.05$), but not in aortic diameter, WSS, or pressure. These results indicate that structural stress is a promising predictor of both the rate and location of aortic growth in uncomplicated TBAD, which supports its use in risk stratification models to identify patients at higher risk of TBAD progression.

## 1. Introduction

Type B aortic dissection (TBAD) is a potentially lethal condition characterized by a tear in the aortic wall distal to the left subclavian artery [1]. This tear allows blood to separate the layers of the aortic wall, resulting in the development of "true" and "false" lumens. Acute TBAD is classified as either uncomplicated or complicated, based upon the presence of organ malperfusion and aortic rupture [2]. The gold standard therapy for acute complicated TBAD is thoracic endovascular aortic repair (TEVAR), which involves the placement of an endoluminal stent graft into the aorta to cover the primary intimal tear, and redirect blood flow into the true lumen. TEVAR has been highly effective in resolving distal malperfusion, and eliminating blood flow into the ruptured false lumen resulting in improved survival in patients with acute complicated TBAD [3, 4]. In contrast, the current 1st line treatment for uncomplicated TBAD is optimal medical therapy (OMT) consisting of aggressive anti-hypertensive therapy combined with surveillance imaging. Although successful in the short term, OMT has poor long-term outcomes with 5-year aortic-intervention free-survival rates of ≤ 50% [5, 6]. Aortic expansion is a major determinant of long-term survival in patients with uncomplicated TBAD, which may be prevented by performing TEVAR in the acute phase. In fact, several studies have demonstrated improved survival with TEVAR compared to OMT for acute uncomplicated TBAD [7-9]. However, TEVAR is not devoid of risk, and has known complications including stroke, paraplegia, acute kidney injury and retrograde type A dissection [9]. Ideally, patients who are at high-risk for rapid aortic expansion could be identified early and undergo TEVAR in the acute phase, and the remaining patients could be treated with OMT and avoid an unnecessary procedure with unnecessary risks.

Previous studies have identified clinical and anatomic predictors of aortic growth in acute uncomplicated TBAD [7, 10-12]. However, maximum aortic diameter remains as the sole predictor of OMT failure that has been consistently identified and validated across multiple studies [11]. Furthermore, aortic diameter is a one-dimensional metric that does not fully capture the three-dimensional (3D) shape features of the aorta. This has led to investigations into alternative non-diameter metrics to predict aortic growth. Our group used statistical shape modeling (SSM) to comprehensively encode the complex 3D TBAD geometry and identify 3D shape features that correlated with aortic growth and failure of OMT in patients with uncomplicated TBAD [13]. We have also identified aortic centerline length and tortuosity as superior predictors of aortic growth compared to maximum aortic diameter [14]. While these findings have improved our predictive capabilities for aortic growth, there is still a lack of knowledge regarding the pathological mechanisms driving aortic growth in TBAD patients.

Biomechanical modeling has shown promise in providing insights into the poorly understood biomechanics of the false lumen wall in TBAD and may provide an improved mechanistic understanding of the factors driving aortic expansion [15, 16]. Computational fluid dynamics (CFD) and four-dimensional (4D) flow magnetic resonance imaging (MRI) have been used to investigate the importance of wall shear stress (WSS), oscillatory shear index, relative residence time, true and false lumen pressure difference, and false lumen perfusion fraction in TBAD [17-21]. However, the role of aortic wall structural stress in TBAD progression has not been sufficiently investigated in previous studies, primarily due to the high computational cost and complexity of fully coupled fluid-structure interaction (FSI) analyses [22-24]. Given that structural wall stress is orders of magnitude greater than hemodynamic WSS, we hypothesize that aortic wall mechanics plays a critical role in driving false lumen expansion and aortic growth in TBAD.

In our initial study using FSI analyses to investigate the impact of structural wall stress on aortic growth in TBAD, we utilized a forward-penalty based approach that enables stress computation without the need for patient-specific material properties or large-deformation simulations [25-27]. This technique enabled the development of an efficient reduced-order FSI workflow that computed structural wall stress in approximately 20 minutes on a desktop PC [28]. Our analysis of 9 acute uncomplicated TBAD patients demonstrated a positive association between structural wall stress and subsequent aortic growth rate. These findings suggest that the spatial distribution of structural wall stress may serve as a promising biomechanical predictor of focal aortic growth.

Building on our previous work, the current study applies the reduced-order FSI analysis to a larger cohort of 30 patients with uncomplicated TBAD. The primary objective of this study is to further investigate the spatial distribution of biomechanical variables, including structural wall stress, WSS, and blood pressure, as potential predictors of the spatial distribution of the aortic growth rate. Additionally, we evaluated differences in the biomechanical variables between patients with stable vs expanding aortic diameter. The identified biomechanical predictors may ultimately be incorporated into a risk stratification model to identify patients at higher risk of TBAD progression and support personalized clinical decision-making.

## 2. Methods

### *2.1. Study design*

This study was approved by the Emory University Institutional Review Board (IRB00109646 and STUDY00002737). The Emory Aortic Database was retrospectively analyzed and 30 patients with acute uncomplicated TBAD initially treated with OMT in Emory Healthcare hospitals from 2008-2024 were identified. Clinical and longitudinal contrast enhanced CT imaging data were collected and analyzed. The workflow of our study is summarized in Figure 1. Structural stress, blood pressure, WSS and aortic diameter (predictor variables) were extracted from the earliest initial CT scans after TBAD diagnosis through the reduced-order FSI analysis. Subsequently, the aortic growth was quantified for each patient by using the earliest CT scan and a follow up CT scan. The aortic growth and predictor variables were subsequently used to conduct the following analyses: (1) mixed-effect regression analyses were performed to examine the relationship between the predictors and the aortic growth rate with consideration of inter-patient variability; (2) mixed-effects logistic regression was employed to perform classification analyses and predict spatial aortic growth based on the predictor variables; (3) group comparison analyses were performed to summarize the spatial distribution of the predictor variables along the descending aorta and to evaluate differences between patients grouped by OMT and aortic growth outcomes.

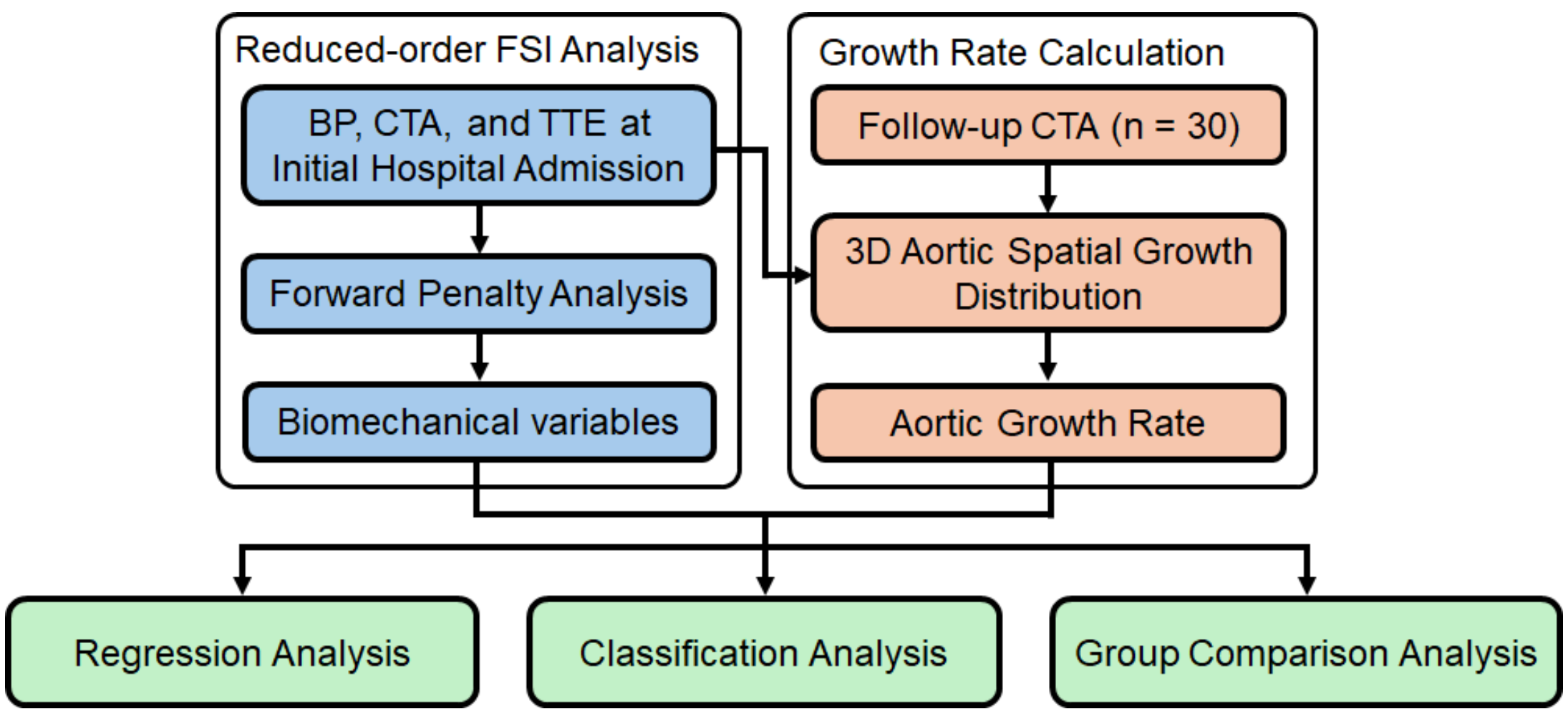

**[Figure 1. Overall analysis workflow: The reduced-order FSI analysis was performed to compute biomechanical variables. Aortic growth rate was computed using both baseline and follow-up aortic geometries. Regression analysis, classification analysis and group comparison analysis were performed to identify biomechanical predictors. CTA: computed**

**tomography angiography; TTE: transthoracic echocardiogram; BP: blood pressure measurement.]**

*2.2 CT image segmentation and meshing*

Three-dimensional (3D) geometries of the thoracoabdominal aorta were segmented for each patient based on baseline CT scans. The segmentations included relevant anatomical features including brachiocephalic artery, left common carotid artery, left subclavian artery, celiac artery, superior mesenteric artery, dissection flap, intimal tears, and intraluminal thrombus. Due to potential motion artifacts in single-phase CT acquisition, all CT images were treated as representing the systolic phase. Depending on the patient's anatomy, each aortic model contained between 7 and 11 branching arteries. Segmentation was performed semi-automatically using 3D Slicer. The "Grow from Seeds" tool was used for the initial segmentation of the aorta, while "Hollow" and "Logical Operators" tools were applied to segment the dissection flap, and the "Threshold" tool was used to segment thrombus when present [29]. Figure 2 (a) to (d) illustrates the CT scan and corresponding segmentations for a representative patient.

To perform reduced-order FSI analysis, computational meshes were constructed from patient-specific geometries based on the following procedure: (1) the watertight surfaces of aortic wall and dissection flap segmentations were remeshed using triangular and quadrilateral elements of approximately equal sizes. For patients with thrombus, triangular elements were generated for the thrombosis surface. The duplicated thrombus surface shared with the false lumen wall was removed, and the nodes located along the outer boundary of the remaining thrombus surface were replaced onto the false lumen surface to ensure nodal connectivity at their interface; (2) to capture the varying thicknesses of the true lumen, false lumen, and dissection flap, we adopted the thickness value from our group's previous experimental work [15] on TBAD tissue specimens.

The true lumen wall was modeled with a uniform thickness of 2 mm, the false lumen wall with 1 mm, and the average thickness measured at 6 spatial locations on the baseline CT scans was used to define the dissection flap thickness; (3) solid domain meshes, including aortic wall, flap, and thrombosis, were generated using C3D8H and C3D10, with an element size of 2mm based on mesh independence tests reported in our previous study [28]. (4) the fluid domain, which represents the blood filled space, was meshed using a combination of 3D hexahedral elements in the boundary layer and tetrahedral elements in the core. Mesh resolution in smaller branches and fenestrations was refined to 25% of the maximum element size. Mesh quality was evaluated and improved based on metrics such as tetrahedral collapse and the Jacobian ratio. Figure 2 (e) to (h) illustrates the solid meshes and fluid meshes through a representative patient in the study.

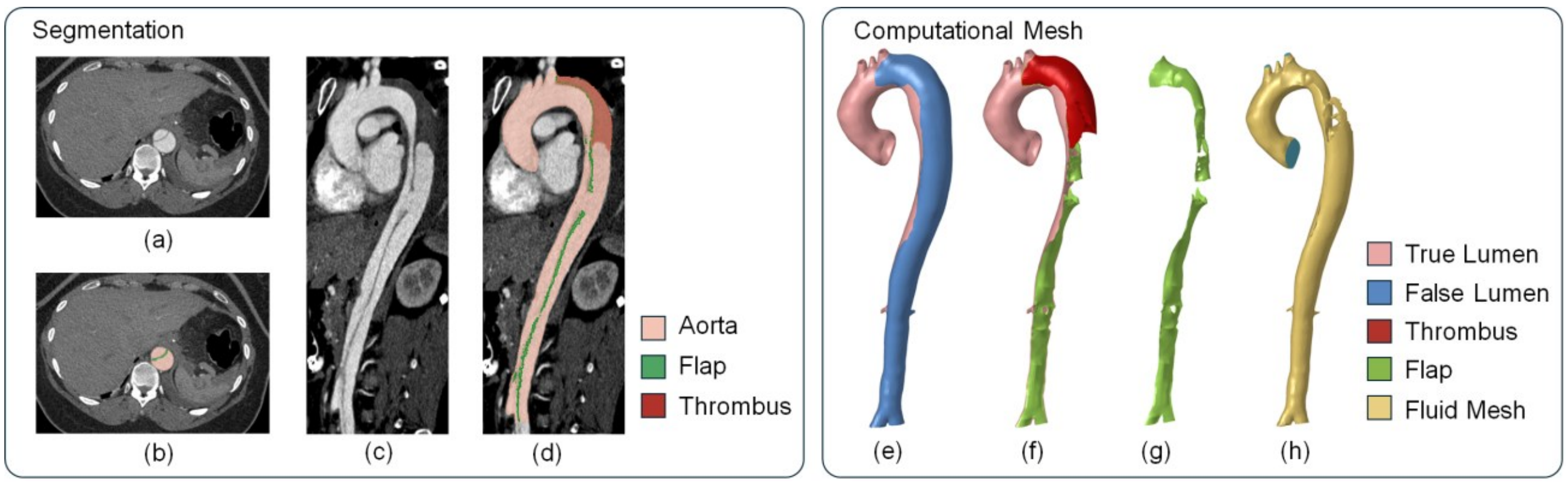

**[Figure 2. CT scan, segmentation, and computational mesh of a representative patient-specific TBAD geometry. (a) axial view of baseline CT scan; (b) axial view of baseline scan with segmentation; (c) sagittal view of baseline CT scan, (d) sagittal view of baseline CT scan with segmentation. (e) Solid mesh of the whole aorta; (f) true lumen wall, flap and thrombus; (g) dissection flap with tears; and (h) fluid domain mesh.]**

*2.3. Reduced-order FSI workflow*

The distributions of structural stress, wall shear stress, and pressure on the aortic wall of baseline CT-derived TBAD geometries were computed using a reduced-order FSI workflow, following our previous work [28]. This reduced-order FSI computes structural stress based on the static determinacy of the aortic wall. An artificial stiff material property is assigned to enforce a penalty treatment, which induces infinitesimal deformations and ensures that the undeformed and deformed geometries remain effectively the same. This approach enables accurate structural stress computation without the need for patient-specific material properties or modeling realistic tissue deformation. Unlike conventional FSI analyses, which require the patient's undeformed geometry and patient-specific material properties, the reduced-order FSI estimates structural stress at the systolic phase while significantly reducing computational cost. The reduced-order FSI computation was carried out in two steps: (1) computational fluid dynamics (CFD) simulations were performed to compute the pressure and WSS distribution at steady-state systolic phase; (2) finite element analysis (FEA) was then performed using the CFD-derived pressure distribution to compute the structural stress distribution at systole.

In the CFD simulations, blood flow within the aorta was modeled with velocity inlet conditions at the aortic root and Windkessel outlet conditions at the branching arteries. The inlet velocity profile followed a 1/7 power-law distribution [30]. For patients P2, P3, P4, P6, P7, P10, P11, P12, P15, P16, P20, P21, P23, and P24, the peak inlet velocity was derived from the peak aortic valve velocity measured by patient-specific echocardiography. For the remaining patients, whom echocardiographic data were unavailable, a population-averaged physiological flow rate during peak systole of 25 L/min was applied at the aortic root inlet [31] with the 1/7 power-law velocity profile. The total vascular resistance for each patient was calculated by dividing the patient-specific measured systolic blood pressure by the corresponding flow rate. Windkessel boundary conditions were applied at each outlet, with resistance distributed proportionally to the cross-sectional area of each branching artery according to Murray's law [32]. To ensure that the simulated systolic blood pressure at the aortic arch matched each patient's clinical measurements, the total vascular resistance was fine-tuned individually. This procedure enabled physiologically accurate simulations of blood flow and aortic wall pressure. The CFD analyses provided both the aortic wall pressure distribution and the WSS for each patient.

To perform FEA, the non-uniform pressure distribution was extracted from the CFD simulations and then mapped onto the solid domain mesh by using 3D interpolation. Leveraging the principle of static determinacy of the aortic wall, an artificial stiff material (Young's modulus $5 \times 10^5$ kPa) was applied to the aortic wall as a penalty treatment, which results in a negligibly small displacement (on the order of $10^{-3}$ millimeter). The Young's modulus for thrombus was set to be $2.5 \times 10^4$ kPa, which maintained the stiffness ratio of 1:20 between aortic wall and thrombus [33]. The maximum principal stress was extracted to represent the structural stress field used in the subsequent analysis. Additionally, circumferential and longitudinal structural stresses were also considered in the regression analysis (see Appendix).

*2.4 Aortic Growth Rate Computation*

For each patient, both baseline and follow-up TBAD geometries were used to calculate the 3D aortic growth rate. The geometries were segmented from the initial baseline CT scan and the most recent follow-up CT scan, and these segmentations were then meshed into 2D surface elements. To ensure consistent geometric representation of TBAD shapes, a mesh parameterization approach [26] was applied to each patient's geometry to generate quadrilateral elements that formed a structured surface mesh. In the structured surface mesh, the aorta was divided into 200 segments (i.e. circumferential layers) along its length, with each circumferential layer consisting of 50 nodes arranged around the circumference of the aorta. The iterative closest point (ICP) algorithm was used to remove rigid body motions and pre-align the meshes [34]. The baseline geometry was then morphed to match the follow-up geometry using Deformetrica software [35, 36], which enables point-to-point correspondence between the baseline and follow-up meshes. The displacement of each node from initial baseline geometry to its corresponding node on the follow-up geometry can be calculated. Based on the displacement field, the logarithmic strain in the circumferential direction was computed, and the aortic growth rate was determined by dividing the strain by the time interval between the baseline and follow-up scans. The steps for the growth rate computation, including pre-alignment, morphing and the resulting growth rate distribution, are illustrated for a representative patient in Figure 3.

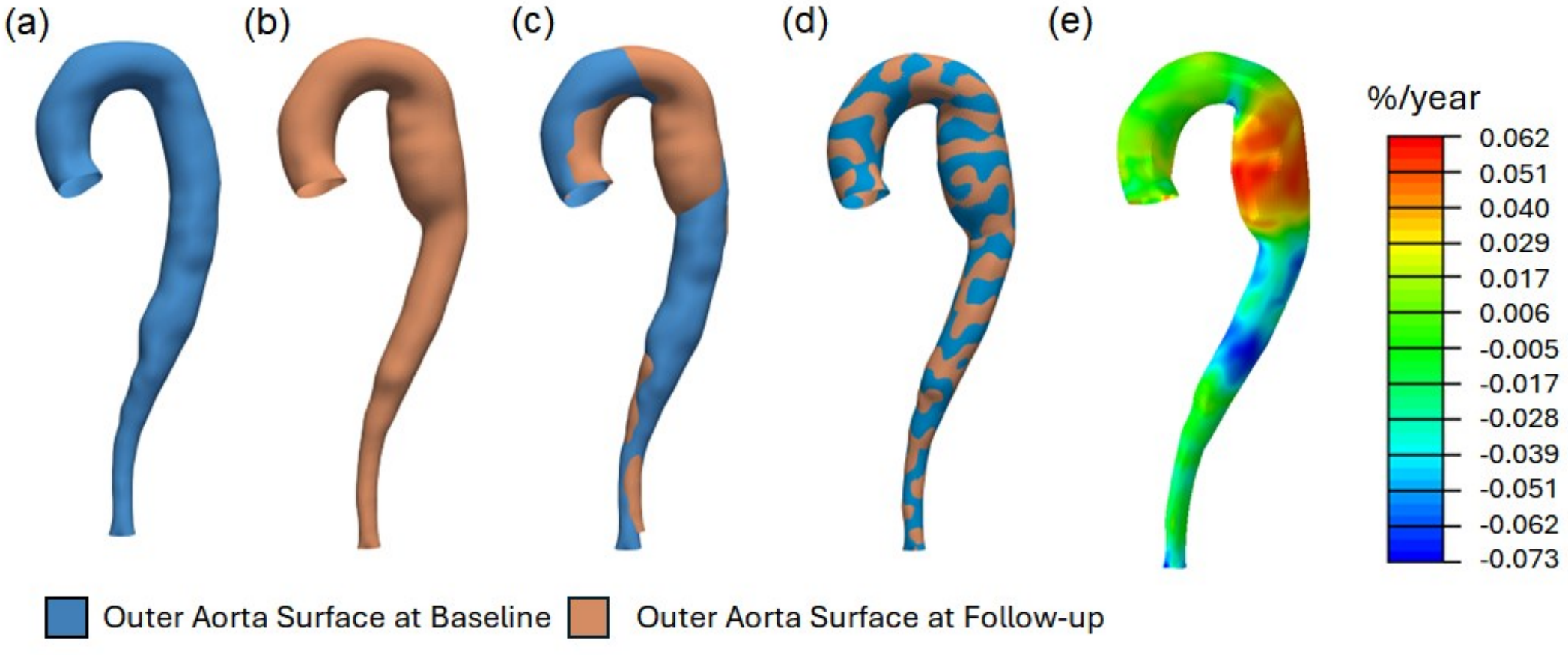

**[Figure 3. Growth rate calculation workflow for a representative patient-specific TBAD geometry. (a) Baseline geometry; (b) follow-up geometry; (c) baseline geometry pre-aligned to follow-up geometry; (d) baseline geometry morphed to the follow-up geometry through Deformetrica software; and (e) computed growth rate (logarithmic strain per year).]**

*2.5. Regression analysis*

To investigate the association between biomechanical predictors and growth rate, linear mixed-effects regression analyses [37] were conducted. In this study, structural stress, WSS, and blood pressure were included as predictors in the analyses. Aortic diameter of the earliest initial geometry was also included as baseline predictor. First, the geometries used for the reduced-order FSI analysis and for aortic growth rate calculation were spatially aligned together to remove rigid body motions. This was achieved by registering the output field (e.g., structural stress) from the reduced-order FSI analysis onto the structured mesh via the same ICP registration procedure used in Section 2.4. To establish point-to-point correspondence between output field and growth rate distributions, for each node on the structured surface mesh, the closest node on the finite element mesh was identified, and the corresponding output field value was assigned to the structured surface mesh, this ensures the output field and growth data share the same 3D coordinates [34]. The sampling method employed in this study is described as follows: The descending aorta, which starts from the left subclavian artery to the aortic bifurcation, is divided into 38 approximately equal-sized regions. Each region contains 3~5 consecutive circumferential layers. The average value of each biomechanical predictor within a region was calculated and used for subsequent analyses. Therefore, for each patient, the spatial distributions of initial aortic diameter, structural stress, WSS, and pressure along the descending aorta were obtained.

Using the linear mixed-effects model, the aortic growth rate can be calculated as:

$$GR_{im} = \beta_0 + \beta_1 \sigma_{im} + b_{0m} + b_{1m} \sigma_{im} + \varepsilon_{im} \quad (1)$$

where $GR_{im}$ represents the growth rate, and $\sigma_{im}$ represents the corresponding predictor (structural stress, WSS, pressure, and initial aortic diameter), $m = 1,2,\dots,30$, denotes patient ID, and $i = 1,2,\dots,1140$, is the data point index. In this study, $GR_{im}$ and $\sigma_{im}$ contains 1140 data points (38 regions each patient and totaling 1140 data points across 30 patients). The linear mixed-effects model decomposes the regression coefficients into fixed-effects (slope $\beta_1$ and intercept $\beta_0$) and random-effects (slope $b_{1m}$ and intercept $b_{0m}$). The fixed effects represent the population-level relationship between the predictors and aortic growth rate across all 30 patients, whereas the random effects capture patient-specific variations in the association. The residual term $\varepsilon_{im}$ accounts for unexplained response at each data point. F-tests were performed to determine the significance of fixed-effect association between the predictors and growth rate, with the null hypothesis that fixed-effect slope $\beta_1 = 0$. The standard deviation and its 95% confidence interval (CI) were calculated for each random-effect term ($b_{1m}$ or $b_{0m}$), and the random effect was considered statistically significant if the CI did not include zero. For each patient, Pearson's correlation coefficient and its corresponding p-value were used to evaluate the association between predictors and aortic growth rate.

*2.6. Classification Analysis*

The linear mixed-effect regression model described in the previous section examined both population-level and patient-level relationships between the predictors and aortic growth rate. In the next step, classification analyses were performed to evaluate whether the biomechanical variables as well as initial aortic diameter can accurately identify regions that experienced aortic growth. To this end, mixed-effect logistic regression [38] was employed to predict aortic diameter growth as a binary outcome variable. To obtain aortic diameter growth distribution for each patient, the average diameters of the 38 regions along the descending aorta were calculated. The differences in aortic diameters between the earliest initial and follow-up TBAD geometries were calculated to quantify aortic diameter growth. A threshold was used to convert the aortic diameter growth (mm) as a continuous variable into binary outcome following previous clinical studies. It has been shown that aortic diameter measurements are prone to errors due to limited image resolution [39, 40], with variations of up to 2 mm typically considered within the measurement uncertainty for CT-based assessments. Therefore, a threshold of 2 mm was employed: regions exhibiting an aortic diameter increase ≥ 2 mm were identified as growth regions, whereas regions with diameter changes below this threshold were identified as non-growth regions.

In the mixed-effects logistic regression model, the growth rate can be modeled as [41]:

$$P\ (Y_{im} = 1) = \frac{1}{1 + \exp\ [-(\beta_0 + \beta_1 \sigma_{im} + b_{0m} + b_{1m}\sigma_{im} + \varepsilon_{im})]} \qquad (2)$$

Where $Y_{im}$ represents the predicted diameter growth, with binary values of 0 and 1: where the value of 1 represents growth, and the value of 0 represents no growth. $P\ (Y_{im} = 1)$ denotes the probability of $Y_{im} = 1$. The predicted probability is converted into binary output by assigning 1 if $P\ (Y_{im} = 1) \geq 0.5$ and 0 if $P\ (Y_{im} = 1) < 0.5$. Similar to the linear mixed-effects model in Section 2.5, $m = 1,2, \dots ,30$ denotes patient ID, and $i = 1,2, \dots ,1140$, indexes the data points. $\sigma_{im}$ represents the predictors, including structural stress, WSS, pressure, and initial aortic diameter. The fixed-effects ($\beta_1$ and $\beta_0$) remained constant across all patients, while the random-effects ($b_{1m}$ and $b_{0m}$) were adjusted for individual patients. F-tests were performed to evaluate the significance of fixed-effect slope, with the null hypothesis that $\beta_1 = 0$.

The threshold of each predictor at the population level was obtained from the fixed-effects of the mixed-effect logistic regression. The threshold can be found by solving for $\sigma$ that leads to $\frac{1}{1+\exp\ [-(\beta_0+\beta_1\sigma)]} = 0.5$, which corresponds to the default cutoff for predicting growth and no growth. Similarly, the predictor threshold of an individual patient can be estimated by solving for $\sigma$ in $\frac{1}{1+\exp\ [-(\beta_0+\beta_1\sigma+ b_{0m}+b_{1m}\sigma)]} = 0.5$. The 95% CIs of the predictors' thresholds were also estimated to characterize their variability at the population-level (from fixed-effects) and at the individual patient level (from both fixed and random effects).

To evaluate the performance of the classification model, receiver operating characteristic (ROC) analyses were conducted. ROC analysis quantifies how well the logistic regression model distinguishes between two classes ("growth" vs "no growth") using the area under the curve (AUC) [42]. A higher AUC indicates better performance: AUC = 1 indicates perfect classification, while AUC = 0.5 indicates the model performance is similar to random guessing.

*2.7. Aortic Growth Group Comparison Analysis*

The regression and classification analyses described in the previous sections focus on the spatial correlation between predictors and aortic growth outcomes. In this section, group comparison analyses were performed to visualize the distributions of various variables along the aortic length in patient groups categorized by aortic growth and OMT outcomes. The distributions of two outcome variables (follow-up aortic diameter and growth rate) and four predictor variables (initial aortic diameter, structural stress, WSS, and pressure) were included. The variables were sampled along the length of descending aorta to generate their corresponding distributions. Specifically, the descending aorta was divided into 15 approximately equal-sized regions, with each region containing 10~12 circumferential layers of the structured mesh. For a variable, the mean ± standard deviation was visualized at each of the 15 regions. The 15 regions can be further grouped into three parts: Part 1 comprises regions 1 to 9, which corresponds to the aortic zone 3, 4 and 5, including the proximal descending aorta from left subclavian to the celiac artery [43]; Part 2 includes region 10, which corresponds to the aortic zone 6, 7, and 8, which covers the celiac artery to the bottom of renal arteries; Part 3 includes regions 11 to 15, which corresponds to the aortic zone 9, which covers the distal descending aorta from the bottom of renal arteries to the iliac arteries.

The cohort of 30 patients was divided into two groups based upon aortic growth and OMT outcomes. Patients with a mean descending aortic diameter increase of ≥ 2mm or a maximum descending aortic growth rate of ≥ 3/year were classified as “enlarging”, and those with mean diameters or growth rates below those thresholds were classified as “stable”. This growth rate threshold (3mm/year) was adopted from previous studies [44-46], which identified 3mm/year as the criterion to categorize the aorta as “stable” or “enlarging”. Patients were also divided into two groups based upon whether they were maintained on OMT alone or required open or endovascular surgical intervention during the study period.

To quantify whether statistical significance exists in the variables between outcome groups, one-tailed two-sample t-tests were performed. The null hypothesis for each variable was formulated as follows: (1) the high-risk groups (diameter growth ⩾ 2 mm, aortic growth rate ⩾ 3 mm/year, and surgery) have smaller initial aortic diameters; (2) smaller follow-up diameters in these groups; (3) slower aortic growth rates in these groups; (4) lower structural stress in these groups; (5) lower pressure in these groups; and (6) higher WSS in these groups. $p < 0.05$ was considered as statistical significance [47] to reject the null hypotheses.

**3. Results**

*3.1. Study sample*

Table 1 lists demographics, comorbidities and anatomic characteristics for all patients. The median duration between the initial and latest follow-up CT scan was 3.47 years with interquartile range of [1.54, 4.79]. The median age of all patients was 55.8 years and 67% were male. 93% had hypertension and 13% had a known diagnosis of a connective tissue disorder. The mean number of visceral branches originating from the true and false lumens were 3.5 and 1.2 respectively. The spatially-averaged aortic growth rate for all patients was 3.53 mm/year.

| | |
|---|---|
| Age at the time of diagnosis (years) | 55.8 [47, 64.75] |
| Sex (% male) | 66.67 |
| Follow-up period (years) | 3.47 [1.54, 4.79] |
| Hypertension (%) | 93.33 |
| Beta-blocker (%) | 46.67 |
| Diabetes mellitus (%) | 16.67 |
| Stroke or transient ischaemic attack (%) | 0 |
| Dyslipidemia (%) | 20.00 |
| End-stage renal disease (%) | 6.67 |
| Smoking (%) | 33.33 |
| Congestive heart failure (%) | 13.33 |
| Connective tissue disorder (%) | 13.33 |
| Number of arteries arising from true lumen | 3.47[2, 5] |
| Number of arteries arising from false lumen | 1.17 [0, 2] |
| False lumen thrombosis (%) | 66.67 |
| Spatially-averaged aortic growth rate (mm/year) | 3.53 [-0.22, 5.88] |
| Systolic brachial blood pressure (mmHg) | 155.73 [127.25, 171.25] |
| Diastolic brachial blood pressure (mmHg) | 80.43 [64.75, 92.5] |

**[Table 1. Patient demographics and clinical variables in the study cohort. The number of arteries includes celiac, superior mesenteric, inferior mesenteric, left renal, and right renal arteries. The median, 25th and 75th percentiles are reported for numerical variables. Percentage is reported for dichotomic variables.]**

*3.2. Contours of structural stress and aortic growth rate*

Spatial distributions of biomechanical variables (structural stress, WSS, and pressure) and aortic growth rate were visualized for 6 representative patients (P12, P14, P15, P17, P22, and P25). Figure 4 shows the structural wall stress, WSS and pressure fields computed from the initial baseline CT, along with the heatmaps of the aortic growth rate displayed on the follow-up geometries of the representative patients. Upon close visual inspection, there are spatial associations between the structural stress distributions and the growth rate heatmaps. In the representative patients P12, P14, P15, P22, and P25 (Figure 4 (a), (b), (c), (e), and (f)), localized regions with high structural wall stress were observed in the descending aorta, which overlaps with regions with significant aortic growth. In patient P17 (Figure 4 (d)), relatively low structural stress was observed along the descending aorta, and the aorta shape remained largely unchanged after 32 months of follow-up, which exhibited a relatively slow growth rate. No clear spatial association with growth rate was observed for either WSS or pressure in the representative patients.

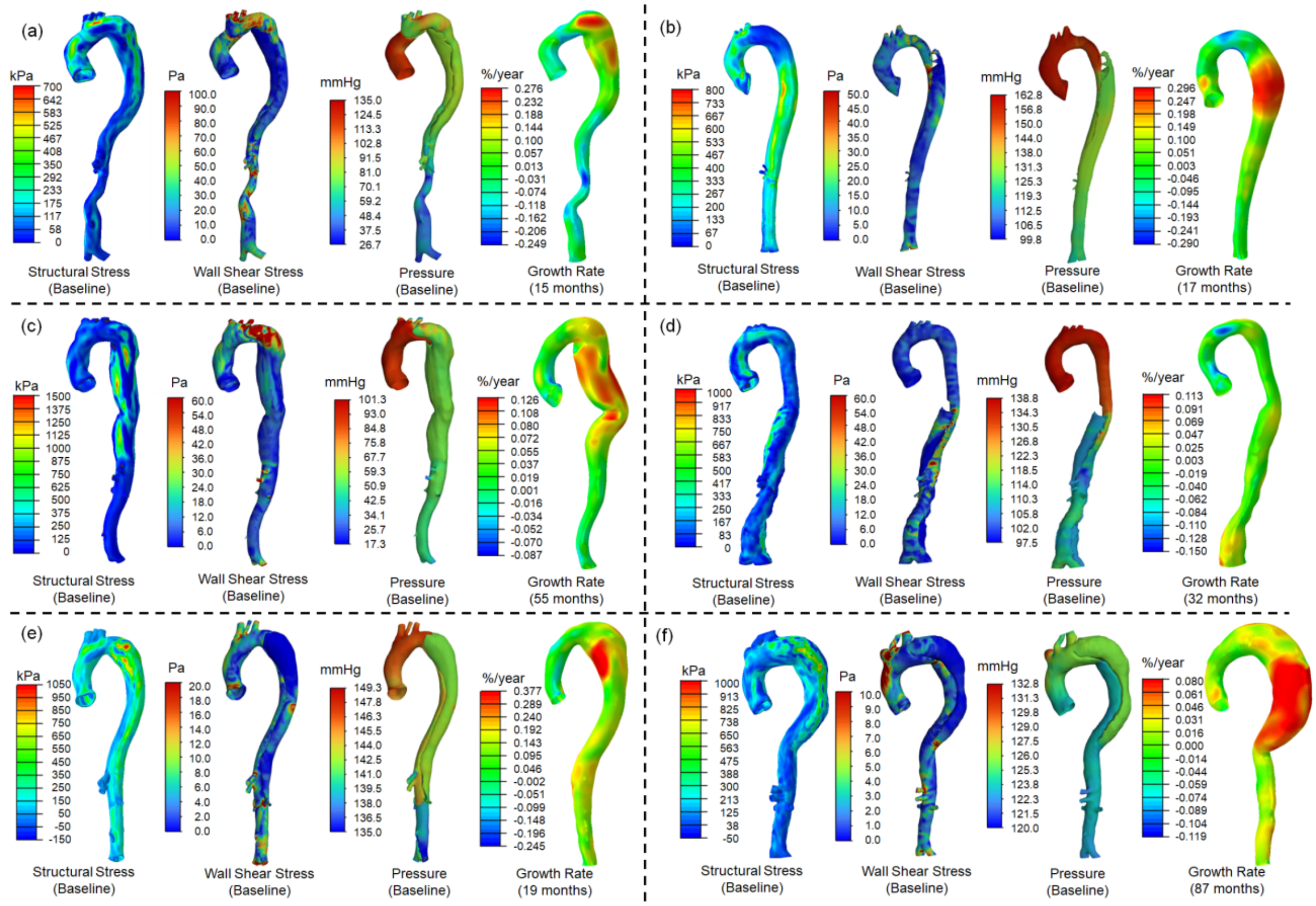

**[Figure 4. Max principal structural stress, WSS and pressure distributions on baseline geometry, along with the corresponding heatmaps of aortic growth rates visualized on follow-up geometry: (a) P12; (b) P14; (c) P15; (d) P17; (e) P22; (f) P25.]**

*3.3. Regression Analyses*

*3.3.1. Fixed-effect associations.*

The fixed-effect results are shown in Figure 5, which illustrates the population-level associations between structural stress, WSS, pressure, initial aortic diameter and the aortic growth rate. While structural stress, pressure, and initial aortic diameter exhibited a positive association, only structural stress showed a statistically significant association (p-value = 0.0003, slope = 0.06%/(year • kPa)) with aortic growth rate. The associations between initial diameter, pressure and aortic growth rate were not statistically significant (initial diameter: p-value = 0.2076; pressure: p-value = 0.4039), and their CI included zero. In contrast, the association between WSS and aortic growth rate was statistically significant (p=0.0227) with a negative slope (-0.11%/(year • Pa)).

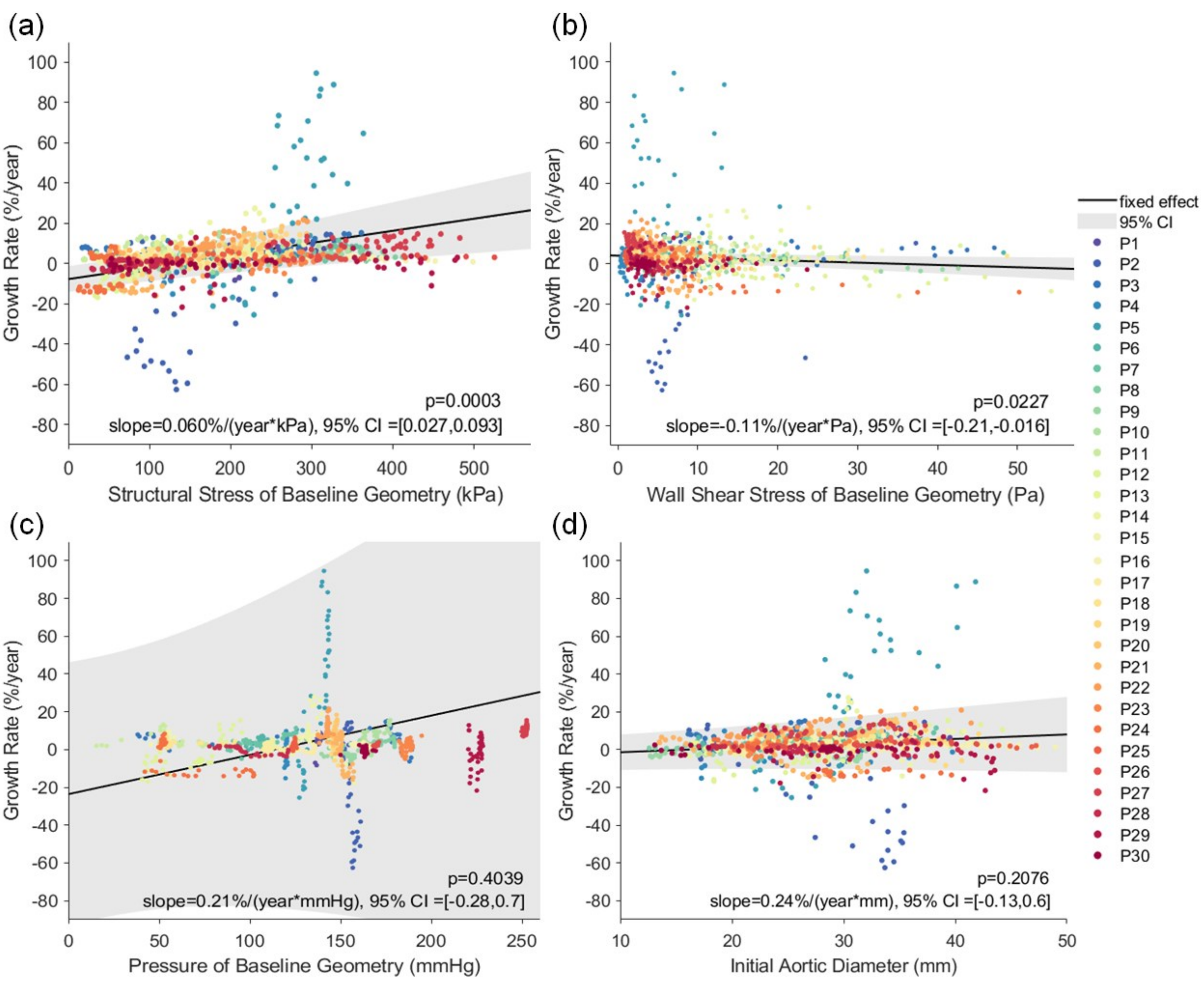

**[Figure 5. (a) Linear mixed-effects regression analysis demonstrates a significant positive fixed-effect association between structural stress of baseline geometry and aortic growth rate (p=0.0003; slope=0.06%/(year*kPa)). (b) the same analysis indicated a significant negative fixed-effect association between WSS of baseline geometry and growth rate (p=0.0227; slope=-0.11%/(year*kPa)). The same analysis did not lead to significant fixed-effect associations using pressure (c) and initial aortic diameter (d) distributions. CI: confidence interval.]**

*3.3.2. Random-effect associations*

Figure 6 summarizes the individual relationships between structural wall stress and growth rate for the 6 representative patients. The regression model predictions incorporated both fixed-effect and random effects. All 6 representative patients exhibit a positive relationship between structural wall stress and growth rate, with statistically significant Pearson correlation coefficients. The data points in Figure 6 are color coded by spatial location, from the proximal to the distal aorta.

Summary statistics of random effects for all patients (n = 30) are as follows: (1) Structural stress and growth rate, 29 out of 30 patients exhibit a positive slope, with 25 of 30 patients showing statistically significant Pearson correlation coefficients ($p < 0.05$). The standard deviations of the random-effects slopes ($b_{1m}$) and intercepts ($b_{0m}$) were estimated to be 0.08%/(year·kPa) with a 95% confidence interval (CI) of [0.07, 0.11] %/(year·kPa), and 17.12% with a 95% CI of [13.63, 23.01] %; (2) WSS and growth rate, all 30 patients exhibit a negative slope, with 22 of 30 patients having statistically significant Pearson correlation coefficients ($p < 0.05$). The standard deviations of the random-effects slopes ($b_{1m}$) and intercepts ($b_{0m}$) were estimated to be 0.04%/(year·Pa) with a 95% confidence interval (CI) of [0.03, 0.05] %/(year·Pa), and 6.72% with a 95% CI of [5.35, 9.03] %; (3) Pressure and growth rate, 21 out of 30 patients exhibits a positive slope, with 21 out of 30 patients have statistically significant Pearson correlation coefficients ($p < 0.05$). The standard deviations of the random-effect slopes ( $b_{1m}$ ) and intercepts ( $b_{0m}$ ) were estimated to be 1.24%/(year·mmHg) with a 95% confidence interval (CI) of [0.98, 1.66] %/(year·mmHg), and 174.26% with a 95% CI of [138.78, 234.26] %; (4) Initial diameter and growth rate, 22 out of 30 patients exhibits a positive slope, with 25 of 30 patients showing statistically significant Pearson correlation coefficients ($p < 0.05$). The standard deviations of the random-effect slopes ($b_{1m}$) and intercepts ($b_{0m}$) were estimated to be 1.24%/(year·mm) with a 95% confidence interval (CI) of [0.98, 1.66] %/(year·mm), and 174.26% with a 95% CI of [138.78, 234.26] %. The individual relationships between WSS, pressure, initial diameter and growth rate are visualized in the Appendix.

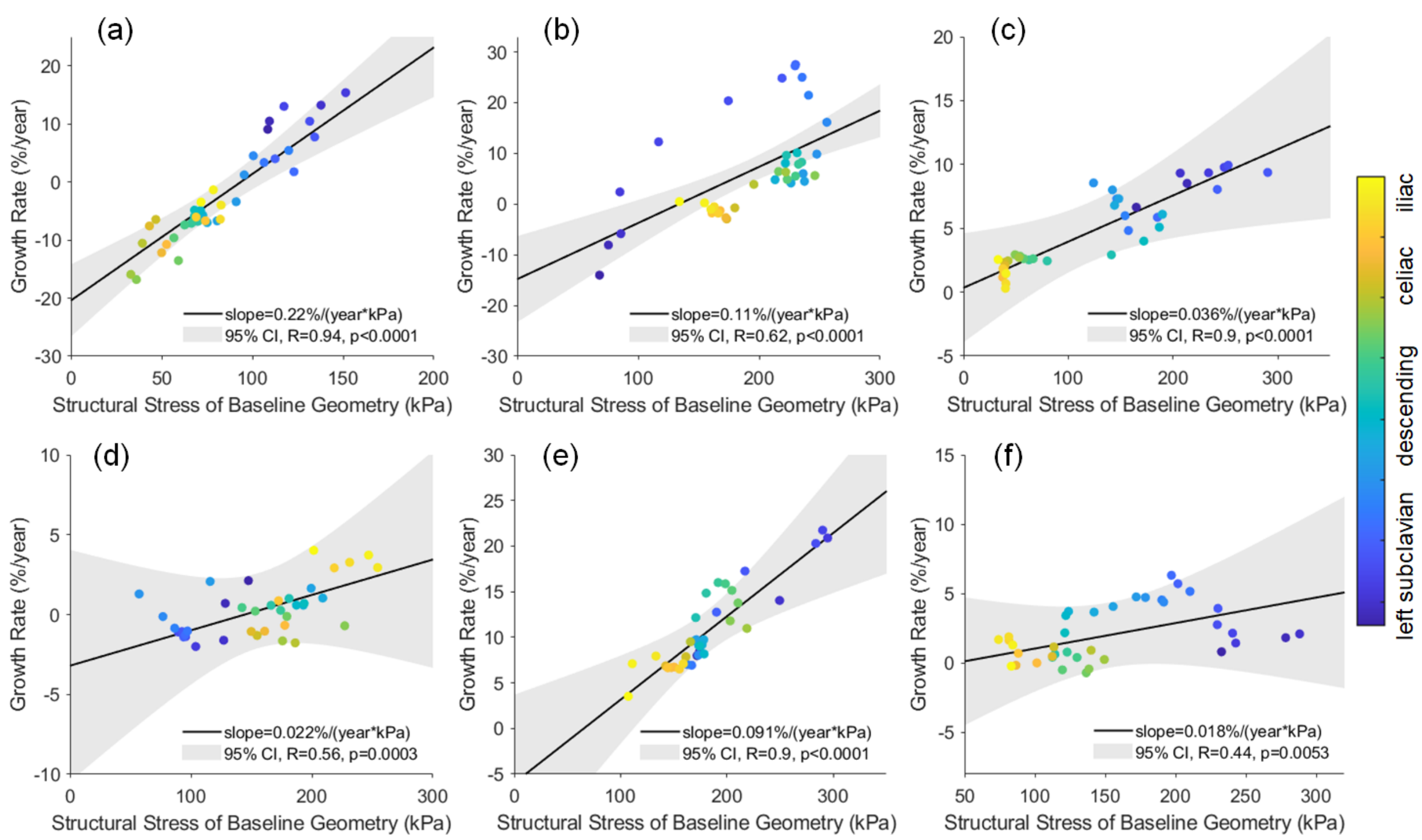

**[Figure 6. Structural stress of baseline geometry and aortic growth rate in 6 representative patients and regression results using linear mixed-effects regression: (a) P12; (b) P14; (c) P15; (d) P17; (e) P22; (f) P25; The data points are color-coded by spatial location from proximal to the distal descending aorta. CI: confidence interval; R, p: Pearson correlation coefficient and its corresponding p-value.]**

### *3.4. Classification Analyses*

#### *3.4.1. Fixed-effect associations*

Figure 7 summarized the fixed-effect results of classification analysis using the predictors: structural stress, WSS, pressure, and initial aortic diameter. The logistic regression model identified the following threshold values of each predictor using the fixed-effects: structural stress = 147.21 kPa; WSS = 9.21 Pa; pressure = 127.92 mmhg; and initial diameter = 26.01mm. Figure 7 (a), (d), (g), and (j) show these thresholds and their 95% CIs for classifying regions with growth vs. no growth. The p-value of fixed-effects slopes for structural stress, WSS, pressure, and initial aortic diameter are $< 0.0001$, 0.0154, 0.0022, $< 0.0001$, respectively. The statistical significance of all predictors indicates that structural stress, WSS, pressure, and initial aortic diameter had substantial discriminative power in the mixed-effect logistic regression. To further demonstrate the discriminative capability of each predictor, Figure 7 (b), (e), (h), and (k) show the ROC curve for the classification based on each individual predictor. Among these predictors, structural stress achieved the highest AUC value of 0.7414, followed by the initial aortic diameter (AUC = 0.6845). WSS exhibited an AUC of 0.5953 and pressure had the lowest AUC of 0.4991. It is worth noting that since WSS led to a negative slope with statistical significance, WSS below the threshold (9.21 Pa) was used to predict growth regions. Figure 7 (c) (f) (i) (l) shows the confusion matrices of the classification results, with structural stress achieving the highest accuracy among all the predictors.

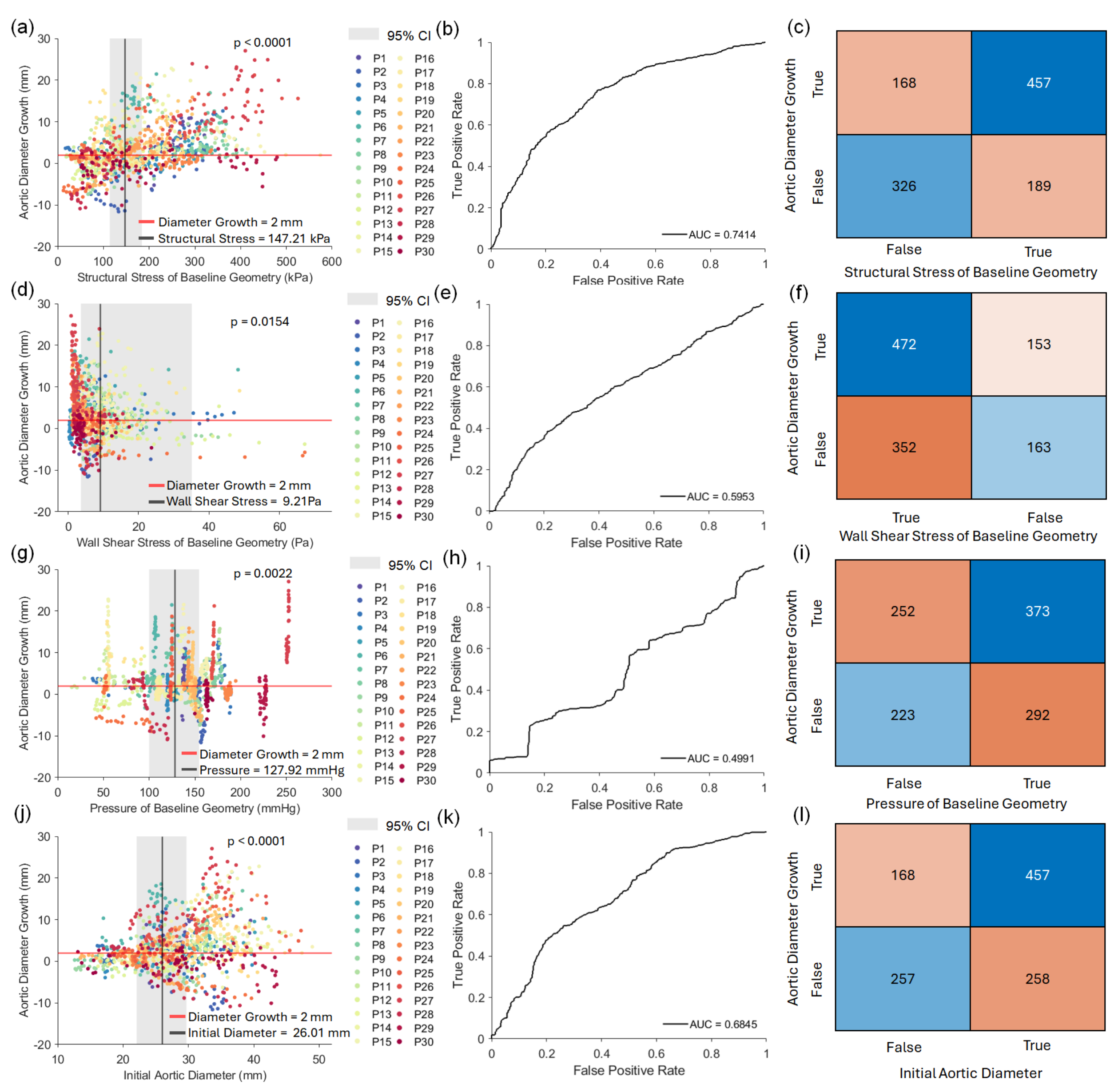

**[Figure 7. Results of the mixed-effects logistic regression analysis. The thresholds: structural stress = 147.21kPa (a); WSS = 9.21Pa (d); pressure = 127.92mmHg (g); and initial aortic diameter = 26.01mm (j). p: p-value of the fixed-effects slope. ROC curve and confusion matrix are presented for structural stress (b, c), WSS (e, f), pressure (h,i), and aortic diameter (k, l).]**

*3.4.2. Random-effect associations.*

Figure 8 presents the logistic regression results using structural stress as the predictor, incorporating both fixed-effects and random-effects for the 6 representative patients. Each subplot displays the classification results of an individual patient. These classification results were consistent with the spatial distributions of structural stress and growth rate in Figure 4. For P15, P22 and P25 (Figure 8 (c), (e), and (f)) substantial diameter changes were observed. Consequently, most data points lie above both the stress and diameter change thresholds and are correctly classified as true positives (purple). For P14, localized diameter changes along the aorta resulted in a more balanced classification (Figure 8 (a)). In contrast, P12 and P17, where most regions of descending aorta exhibited minimal growth and generally low structural stress levels, were predominantly correctly classified as true negatives (blue), as shown in Figure 8 (b) and (d).

Summary statistics of random effects for all patients (n = 30) are as follows: (1) Structural stress and growth rate, 29 out of 30 patients exhibit a positive slope. The standard deviations of the random-effects slopes ($b_{1m}$) and intercepts ($b_{0m}$) were estimated to be 0.01/kPa with a 95% confidence interval (CI) of [0.01, 0.02] /kPa, and 2.42 with a 95% CI of [1.93, 3.25]; (2) WSS and growth rate, 23 of 30 patients exhibit a negative slope. The standard deviations of the random-effects slopes ($b_{1m}$) and intercepts ($b_{0m}$) were estimated to be 0.16/Pa with a 95% confidence interval (CI) of [0.13, 0.21] /Pa, and 1.68 with a 95% CI of [1.34, 2.26]; (3) Pressure and growth rate, 25 out of 30 patients exhibits a positive slope. The standard deviations of the random-effect slopes ($b_{1m}$) and intercepts ($b_{0m}$) were estimated to be 0.08/mmHg with a 95% confidence interval (CI) of [0.07, 0.11] /mmHg, and 10.28 with a 95% CI of [8.19, 13.83]; (4) Initial diameter and growth rate, 26 out of 30 patients exhibits a positive slope. The standard deviations of the random-effect slopes ($b_{1m}$) and intercepts ($b_{0m}$) were estimated to be 0.20/mm with a 95% confidence interval (CI) of [0.16, 0.27] /mm, and 5.70 with a 95% CI of [4.54, 7.66].

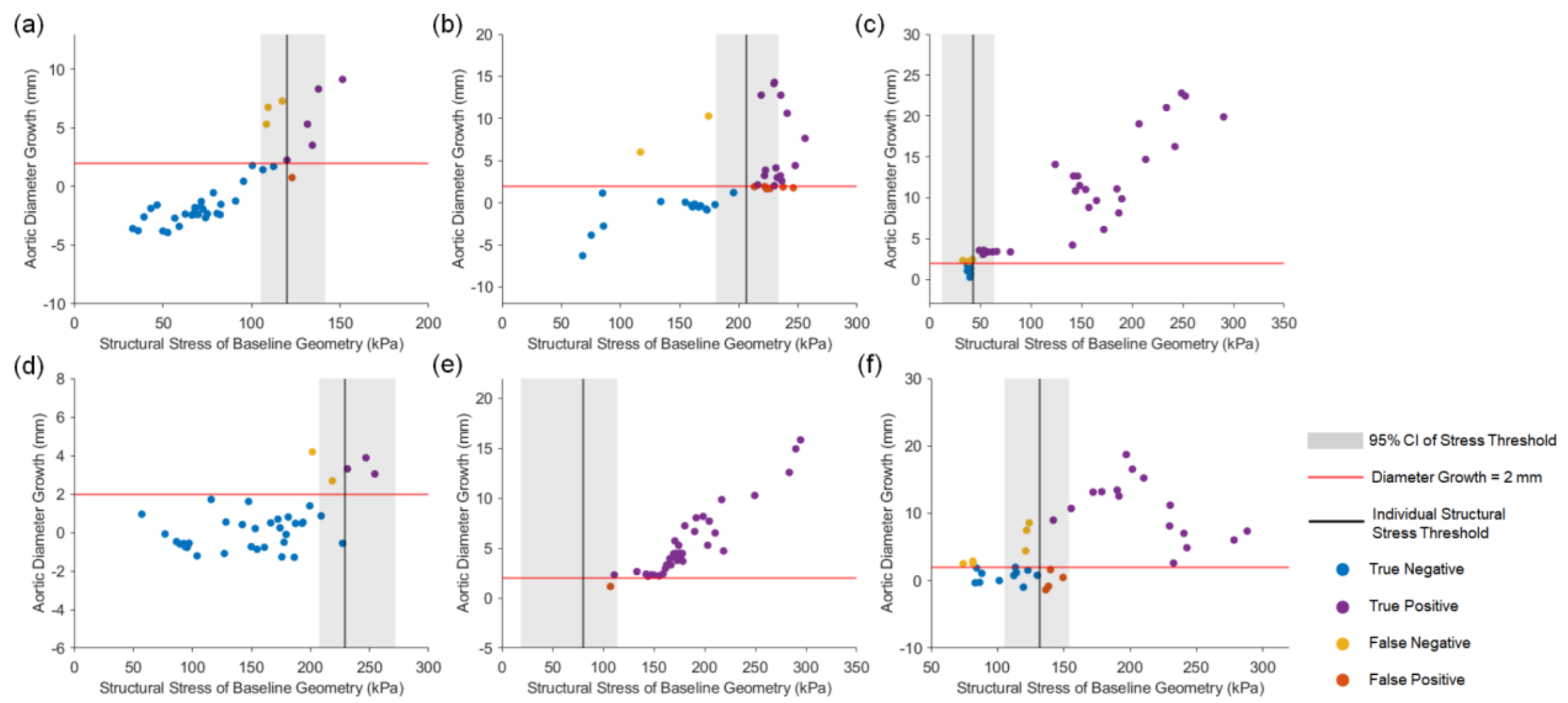

**[Figure 8. Structural stress of baseline geometry and aortic growth rate in 6 representative patients and classification results using mixed-effects logistic regression. (a) P12, (b) P14, (c) P15, (d) P17, (e) P22, and (f) P25. Blue: true negative, purple: true positive, yellow: false negative, and red: false positive. Red horizontal line is the diameter change threshold = 2mm, and the black vertical line is the patient-specific structural stress threshold obtained from the classification model. CI: confidence interval.]**

*3.5. Group comparison analysis outcome*

Figure 9 presents the distributions of a predictor variable (initial aortic diameter) and two outcome variables (follow-up diameter and aortic growth rate) along the normalized length of the descending aorta, for groups categorized by the aortic growth and OMT outcomes. The high-risk groups are highlighted in red, defined as patients with the following criteria: 1) an average descending aortic diameter growth ≥ 2mm (n = 11); 2) a maximum descending aorta growth rate ≥ 3mm/year (n = 10) or; 3) open or endovascular surgical intervention during the study period (n = 14). The initial aortic diameter did not lead to any statistically significant p-values across the aortic zones (Part 1: zones 3, 4, and 5; Part 2: zones 6, 7, and 8; Part 3: zone 9). The follow-up aortic diameter significantly differentiated all aortic zones in patients grouped by average diameter growth, as shown in Figure 9 (b). It also differentiated aortic zones 3, 4, and 5 between OMT and surgery groups with statistical significance. The aortic growth rate significantly differentiated all aortic zones in patients grouped by average diameter growth, as shown in Figure 9 (c). It also significantly separated aortic zones 6, 7, 8, and 9 in patients grouped by average aortic growth rate, and also differentiated zones 3, 4, and 5 between the OMT and surgery groups with statistical significance. In addition, the figures show that the proximal descending aorta generally has a larger diameter than the distal zones, for both initial and follow-up geometries.

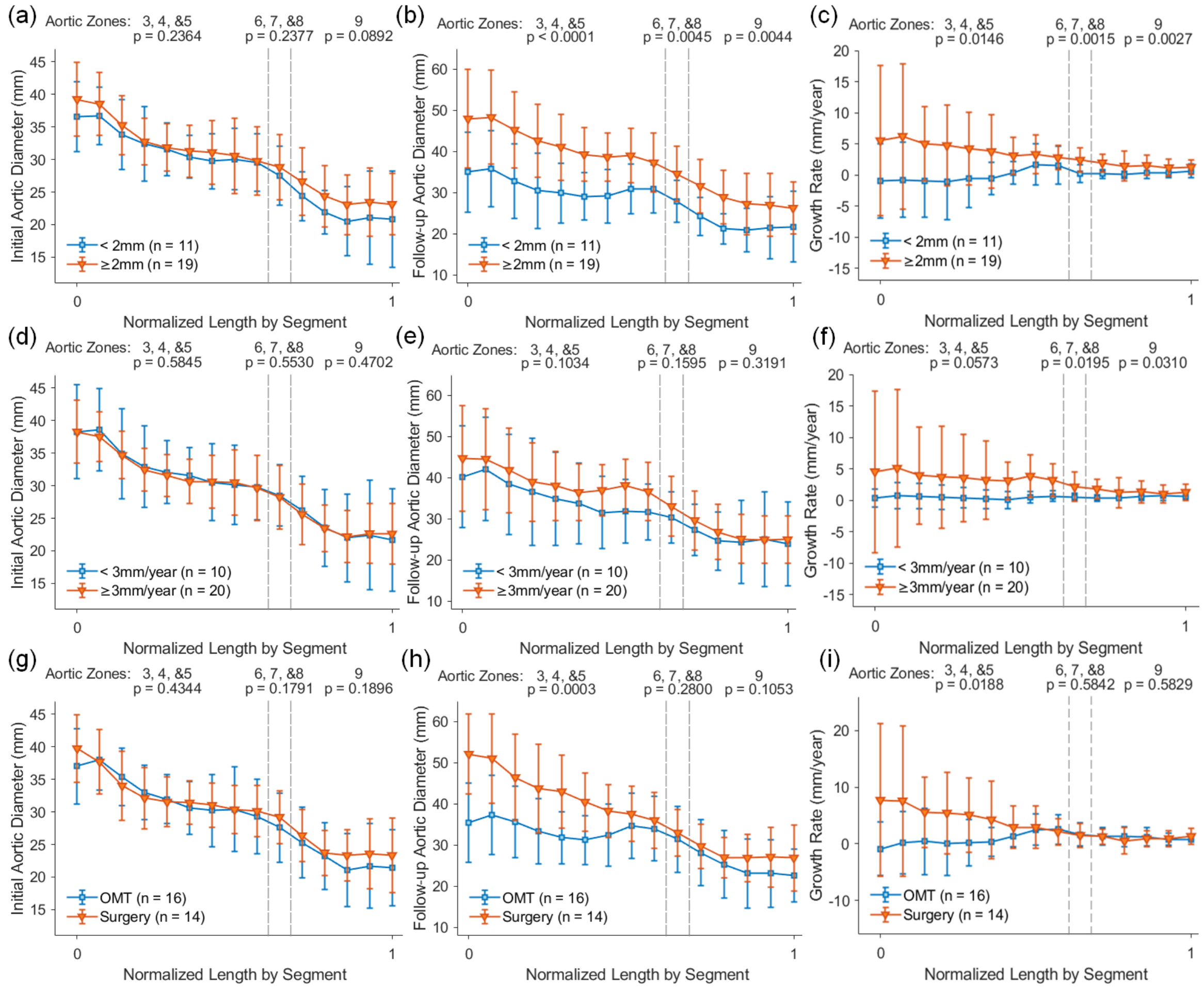

**[Figure 9. Distributions of initial aortic diameter, follow-up aortic diameter, and aortic diameter growth rate along the normalized descending aorta length for patients' groups categorized by OMT and aortic growth outcomes. (a), (b) and (c): initial aortic diameter, follow-up aortic diameter, and growth rate distribution in diameter growth ≥ 2mm and < 2mm groups; (d), (e) and (f): initial aortic diameter, follow-up aortic diameter, and growth rate distribution in diameter growth rate ≥ 3mm/year and < 3mm/year groups; (g), (h) and (i): initial aortic diameter, follow-up aortic diameter, and growth rate distribution in OMT and surgery groups.]**

Figure 10 presents the distribution of biomechanical variables (structural stress, WSS, and pressure) along the normalized length of the descending aorta, for groups categorized by the aortic growth and OMT outcomes. The structural stress significantly distinguished all aortic zones in patients grouped by average diameter growth, as shown in Figure 10 (a). It also statistically significantly differentiated aortic zones 6, 7, and 8 between OMT and surgery groups. The WSS and pressure did not lead to any statistically significant p-values across the aortic zones from zone 3 to zone 9. In addition, the figures show that structural stress of high-risk groups is generally higher than the low-risk groups across the entire descending aorta. WSS distributions are similar between high and low-risk groups, with regions 4, 5, 10, 11, and 14 exhibiting relatively higher WSS magnitudes and greater variability compared to other regions. Pressure distributions along the descending aorta are similar between high and low risk groups, overall, the pressure gradually decreases from the proximal to distal descending aorta.

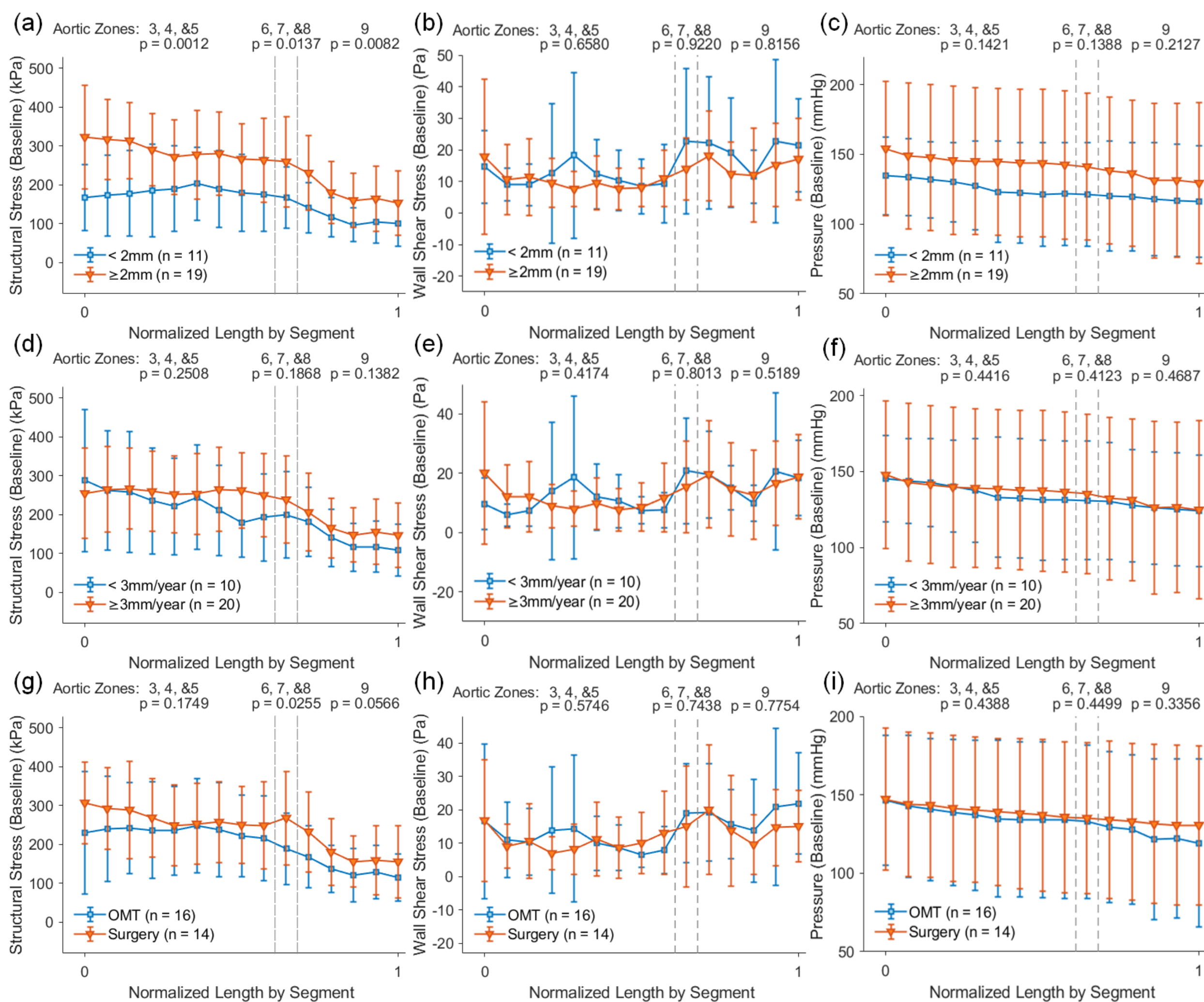

**[Figure 10. Distributions of structural stress, WSS, and pressure along the normalized descending aorta length for patients' groups categorized by OMT and aortic growth outcomes. (a), (b) and (c): Structural stress, WSS, and pressure distribution in diameter growth ≥ 2mm and < 2mm groups; (d), (e) and (f): Structural stress, WSS, and pressure distribution in diameter growth rate ≥ 3mm/year and < 3mm/year groups; (g), (h) and (i): Structural stress, WSS, and pressure distribution in OMT and surgery groups.]**

## 4. Discussion

In the present study, the relationship between biomechanical variables and aortic growth outcomes was investigated by analyzing a cohort of 30 patients with acute uncomplicated TBAD.

The biomechanical predictors (structural stress, WSS, and pressure) were computed through a reduced-order FSI analysis based on the workflow proposed in our previous study [28]. The initial aortic diameter was also included as a baseline for comparison. Compared with traditional fully-coupled FSI simulations, which can require several days to complete [24, 48], the forward penalty based reduced order FSI simulations do not require patient specific material properties or realistic tissue deformations [21, 26, 27], and therefore reduces computational time to approximately 20 minutes on a standard desktop PC. Results from the regression analyses, classification analyses, and group comparison analyses demonstrated that structural stress exhibits a statistically significant positive spatial relationship with the aortic growth rate, achieves the highest AUC in classifying regions with aortic growth, and shows a statistically significant difference between high-risk and low-risk groups. WSS was also found to have a statistically significant negative spatial relationship with the aortic growth rate. These findings suggest that structural stress and WSS may serve as promising biomechanical predictors of aortic growth in TBAD. Furthermore, the classification analysis indicates that a structural stress threshold may exist, beyond which significant aortic growth occurs. This observation could further inform the development of predictive tools to identify spatial locations at high risk for aortic enlargement. However, this concept and the corresponding threshold requires further validation in future studies with a larger patient cohort.

In the present work, the regression analyses showed that WSS has a statistically significantly negative association with aortic growth rate. Generally, lower false lumen WSS reflects lower flow velocity, which may indicate a smaller pressure difference between the true and false lumens. Because true lumen pressure is typically higher, lower WSS would bring false lumen pressure closer to true lumen pressure, which leads to higher structural stress and, consequently, a higher aortic growth rate. However, the exact mechanism may be influenced by patient-specific anatomy and the location of the primary intimal tear. Overall, while WSS appears to correlate with aortic growth, its role in TBAD progression requires further investigation.

The blood pressure distribution showed a positive association with aortic growth rate; however, the F-test of fixed-effect slope indicated that this relationship was not statistically significant. This may be explained by the fact that pressure exhibits greater variability across patients than within an individual patient, which results in limited sensitivity to spatial differences. Moreover, pressure demonstrated the lowest performance in the classification analyses and failed to distinguish between high-risk and low-risk patients. These findings suggest that pressure alone may not serve as a direct predictor of aortic growth. Nevertheless, as discussed previously, the hemodynamic properties of TBAD can influence false lumen pressure and thus thrombosis formation. Therefore, the pressure difference between the true and false lumen, rather than the absolute pressure, may serve as a potential predictor of aortic expansion in TBAD, consistent with previous findings in the literature [49].

Although the relationship between aortic diameter and growth rate was not statistically significant, the classification analysis produced an AUC of 0.6845, which indicates that initial diameter has moderate discriminatory performance [42]; however, this performance level is not clinically sufficient. This finding aligns with clinical observations that diameter remains the only clinically reproducible predictor of TBAD outcomes [50]. According to Laplace's law [51], a larger initial diameter generally leads to a higher structural stress on the aortic wall. However, because structural stress is influenced by multiple factors beyond diameter alone, relying solely on the initial diameter as a predictor may be insufficient.

This study has several assumptions and limitations. First, steady-state condition was assumed in the CFD analyses, and structural stress is computed from reduced-order FSI analyses. This modeling scheme does not capture the dynamic motion of the dissection flap throughout the cardiac cycle or provide time-resolved hemodynamic insights. Nevertheless, the reduced-order FSI analysis significantly reduces the computation time compared to conventional fully-coupled FSI, and therefore allows the inclusion of a larger patient cohort to investigate the relationship between biomechanical predictors and aortic growth. Second, patient-specific TBAD wall thickness may be difficult to obtain from CT images, the TBAD geometries were constructed using constant uniform wall thickness based on our group's experimental work [15]: 2 mm for the true lumen wall, and 1 mm for the false lumen wall. It is known that wall thickness varies both among patients and across different regions within the same patient. Third, our current analysis utilized follow-up CT scans acquired at nonuniform time intervals. Previous studies [52] have shown that the TBAD aortic growth is nonlinear and differs between the acute and chronic phases, with rapid growth typically occurring during the acute phase and slower growth during the chronic phase. However, in this study, the growth rate was calculated by dividing the observed growth by the follow-up time interval. Future studies could benefit from using CT images obtained at a standardized interval or from interpolating geometries at fixed time points using TBAD aortic growth prediction models.

Future studies could investigate the following: (1) Incorporating additional clinical variables, such as smoking history and genetic factors, to improve predictive capability using multivariate mixed-effects regression. Including additional anatomical features of the aorta, such as curvature and centerline metrics [14, 53], in regression and classification analyses could also provide further insight into aortic growth; (2) Expanding the cohort size to enhance the performance of predictive models; and (3) Integrating biomechanical and shape features with machine learning approaches to develop more powerful predictive tools that could support personalized treatment strategies in clinical practice [54, 55].

## 5. Conclusions

This study employed reduced-order fluid-structure interaction (FSI) analysis to compute the distribution of structural stress, wall shear stress (WSS), and pressure in patients with acute uncomplicated type B aortic dissection (TBAD). These biomechanical variables were evaluated for their relationships with aortic growth outcomes using regression analyses, classification analyses, and group comparison analyses. Initial aortic diameter was also included for comparison. Linear mixed-effects regression analyses revealed significant associations between aortic growth rate and both structural stress and WSS, whereas pressure and initial diameter showed no significant relationships. Among all predictors, structural stress demonstrated the best performance in the classification analysis and was the only variable that significantly differentiated specific aortic zones between high-risk and low-risk patient groups in the group comparison analysis. Although these findings suggest that structural stress may serve as a promising predictor for TBAD risk stratification, further cross-validation in a larger cohort is warranted to confirm its predictive capability.

## 6. Acknowledgements

This study is supported in part by the National Institutes of Health (R01HL155537).

The authors thank Kinza Johnson, Marina Dominguez, and Laurel Alonzo for significant contributions to image segmentation. The authors also thank Kristina Porte for significant contributions to clinical data acquisition.

## 7. Declaration of Interest Statement

The authors declare that they have no known competing financial interests or personal relationships that could have appeared to influence the work reported in this paper.

## 8. Appendix

*8.1. Regression analysis of directional structural stress and growth rate*

*8.1.1. Fixed-effect associations of directional structural stress and growth rate.*

Figure A1 shows the population-level associations between structural stress and growth rate in both circumferential and longitudinal directions. All fixed effects in Figure A1 showed significantly positive associations: (1) Circumferential structural stress and circumferential growth rate ($p = 0.0001$, slope = 0.062/(year • kPa)); (2) Circumferential structural stress and longitudinal growth rate ($p = 0.0001$, slope = 0.018/(year • kPa)); (3) Longitudinal structural stress and circumferential growth rate ($p = 0.0002$, slope = 0.10/(year • kPa)); (4) Longitudinal structural stress and longitudinal growth rate ($p = 0.0082$, slope = 0.050/(year • kPa));

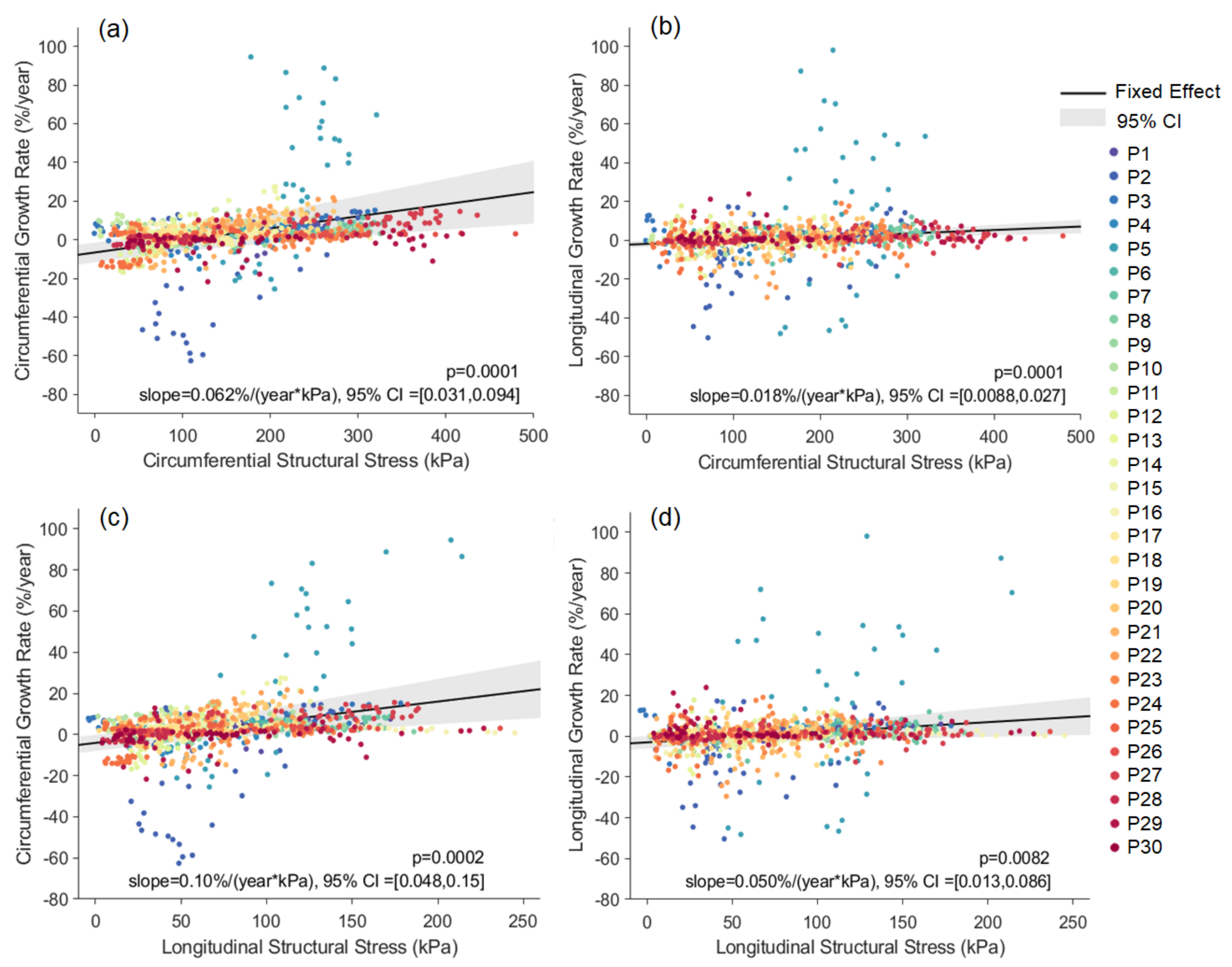

**[Figure A1. Linear mixed-effects regression analysis demonstrates significant positive fixed-effect associations between (a) circumferential structural stress and circumferential growth rate (p=0.0001; slope=0.062%/(year*kPa), (b) circumferential structural stress and longitudinal growth rate (p=0.0001; slope=0.018%/(year*kPa), (c) longitudinal structural stress and circumferential growth rate (p=0.0002; slope=0.1%/(year*kPa), and (d) longitudinal structural stress and longitudinal growth rate (p=0.0082; slope=0.05%/(year*kPa) CI: confidence interval.]**

*8.1.2. Random-effect associations of directional structural stress and growth rate.*

Summary statistics of random effects for all patients (n = 30) are as follows: (1) Circumferential structural stress and circumferential growth rate: 29 out of 30 patients exhibit a positive slope, with 26 of 30 patients showing statistically significant Pearson correlation coefficients ($p < 0.05$). The standard deviations of the random-effects slopes ($b_{1m}$) and intercepts ($b_{0m}$) were estimated to be 0.08%/(year·kPa) with a 95% confidence interval (CI) of [0.06, 0.11] %/(year·kPa), and 12.99% with a 95% CI of [10.34, 17.46] % ; (2) Circumferential structural stress and longitudinal growth rate: all of 30 patients exhibit a positive slope, with 11 of 30 patients showing statistically significant Pearson correlation coefficients ($p < 0.05$). The standard deviations of the random-effects slopes ($b_{1m}$) and intercepts ($b_{0m}$) were estimated to be 0.004%/(year·kPa) with a 95% confidence interval (CI) of [0.003, 0.005] %/(year·kPa), and 1.33 with a 95% CI of [1.06, 1.79] % ; (3) Longitudinal structural stress and circumferential growth rate, 28 out of 30 patients exhibit a positive slope, with 23 of 30 patients showing statistically significant Pearson correlation coefficients ($p < 0.05$). The standard deviations of the random-effect slopes ($b_{1m}$) and intercepts ($b_{0m}$) were estimated to be 0.13%/(year·kPa) with a 95% confidence interval (CI) of [0.11, 0.18] %/(year·kPa), and 10.21% with a 95% CI of [8.13, 13.72] %; (4) Longitudinal structural stress and longitudinal growth rate, 28 out of 30 patients exhibit a positive slope, with 15 of 30 patients showing statistically significant Pearson correlation coefficients ($p < 0.05$). The standard deviations of the random-effect slopes ($b_{1m}$) and intercepts ($b_{0m}$) were estimated to be 0.07%/(year·kPa) with a 95% confidence interval (CI) of [0.06, 0.09] %/(year·kPa), and 6.51% with a 95% CI of [5.18, 8.75] %.

*8.2. Regression analyses using wall shear stress (WSS), pressure and initial diameter*

Figure A2 summarizes the individual correlations between WSS distributions and spatial growth rate distribution for representative patients (P12, P14, P15, P17, P22, and P25). 3 out of 6 patients have a statistically significant Pearson correlation coefficients and all of them have a negative slope.

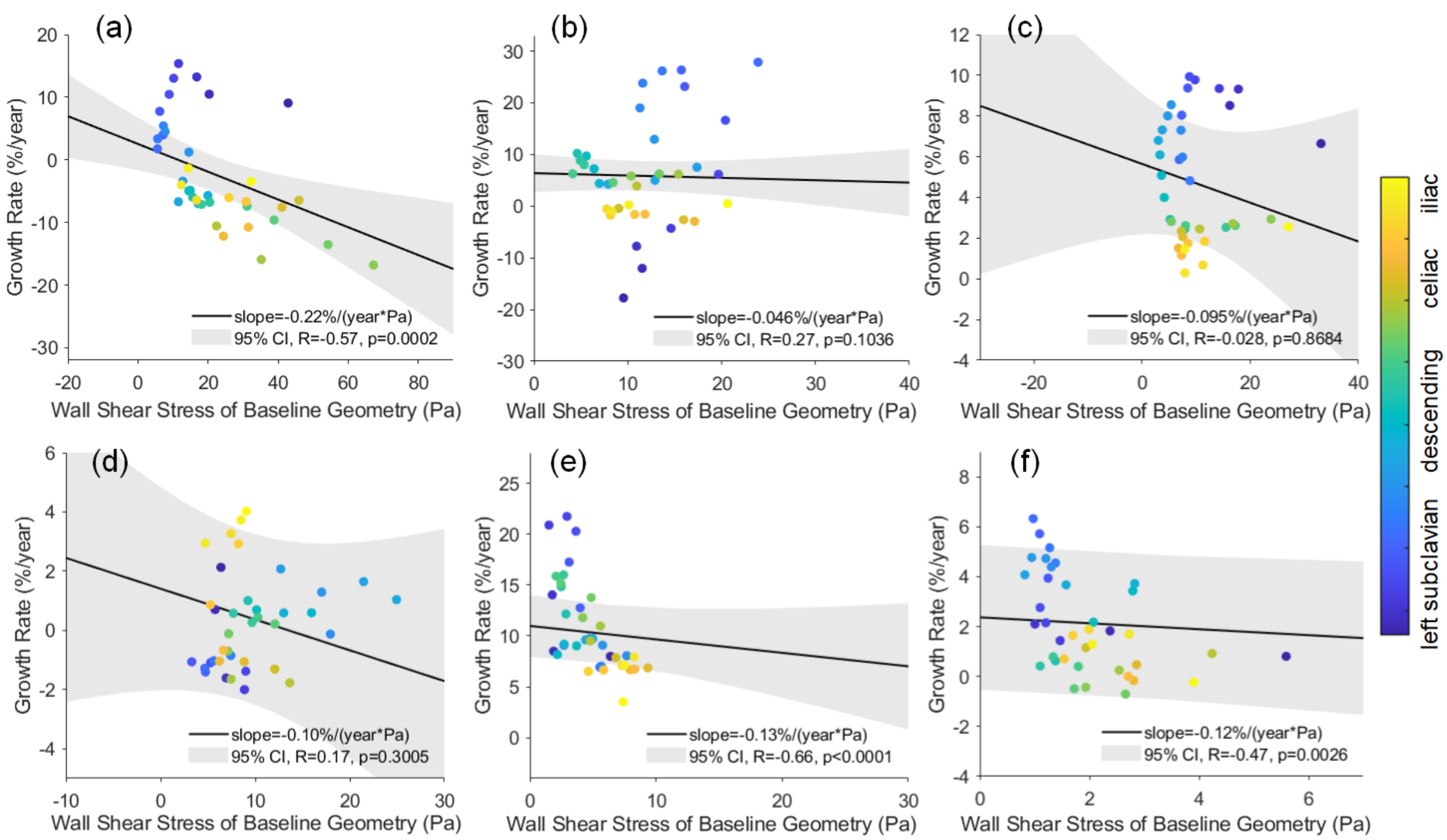

**[Figure A2. WSS distribution of baseline geometry and aortic growth rate in 6 representative patients. (a) P12; (b) P14; (c) P15; (d) P17; (e) P22; (f) P25. Data points are color-coded by spatial location from the proximal to the distal aorta. The regression lines were obtained by using both fixed- and random- effects in the linear mixed-effects model. CI: confidence interval; R, p: Pearson correlation coefficient and its corresponding p-value.]**

Figure A3 summarizes the individual correlations between pressure distributions and spatial growth rate distribution for representative patients (P12, P14, P15, P17, P22, and P25). 5 out of 6 patients have a statistically significant Pearson correlation coefficients, and 2 out of 6 patients have a negative slope.

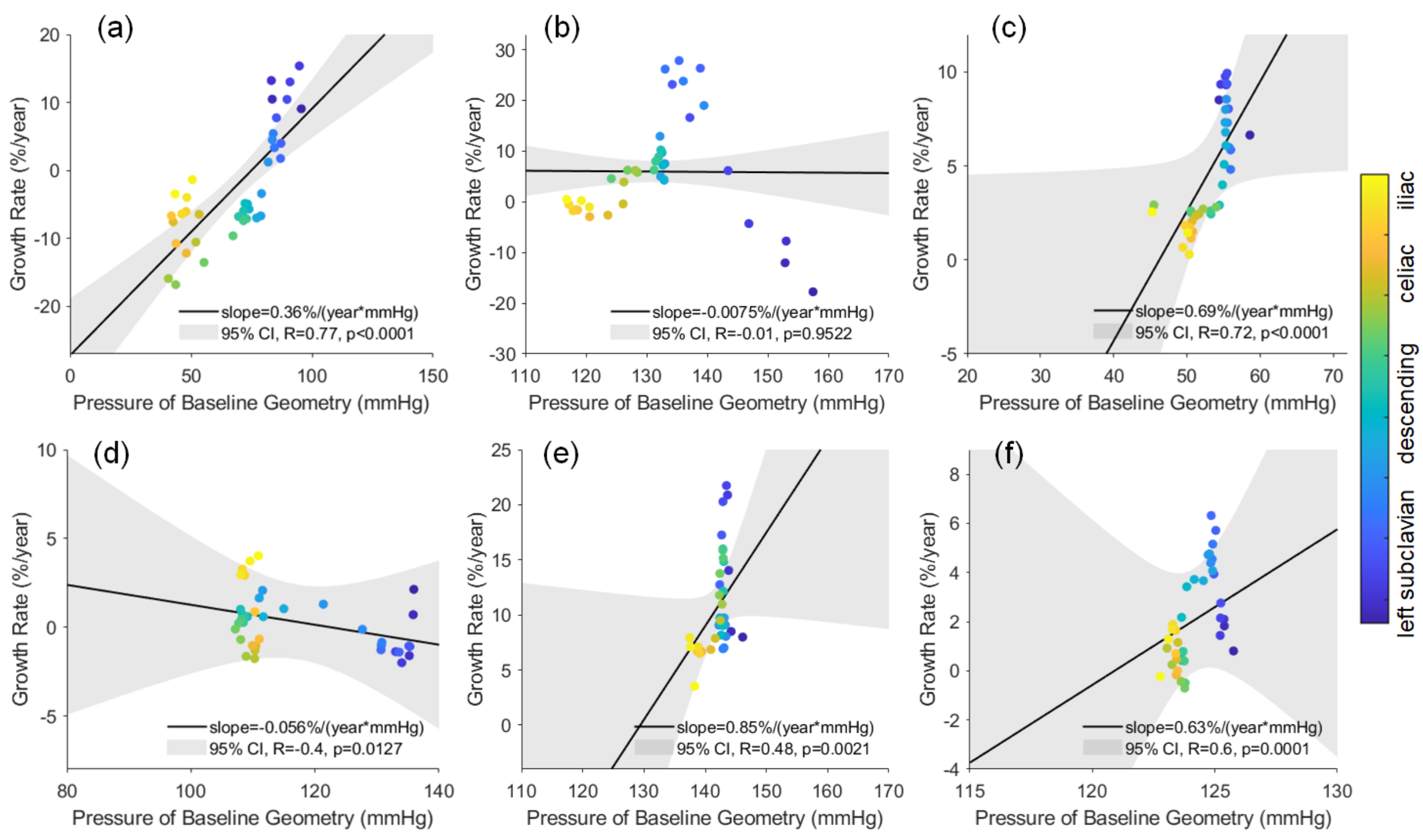

**[Figure A3. Pressure distribution of baseline geometry and aortic growth rate in 6 representative patients. (a) P12; (b) P14; (c) P15; (d) P17; (e) P22; (f) P25. Data points are color-coded by spatial location from the proximal to the distal aorta. The regression lines were obtained by using both fixed- and random-effects in the linear mixed-effects model. CI: confidence interval; R, p: Pearson correlation coefficient and p-value.]**

Figure A4 summarizes the individual correlations between initial diameter and 3D spatial growth rate distribution for representative patients (P12, P14, P15, P17, P22, and P25). 5 out of 6 patients have a statistically significant Pearson correlation coefficients, and all of 6 patients have a positive slope.

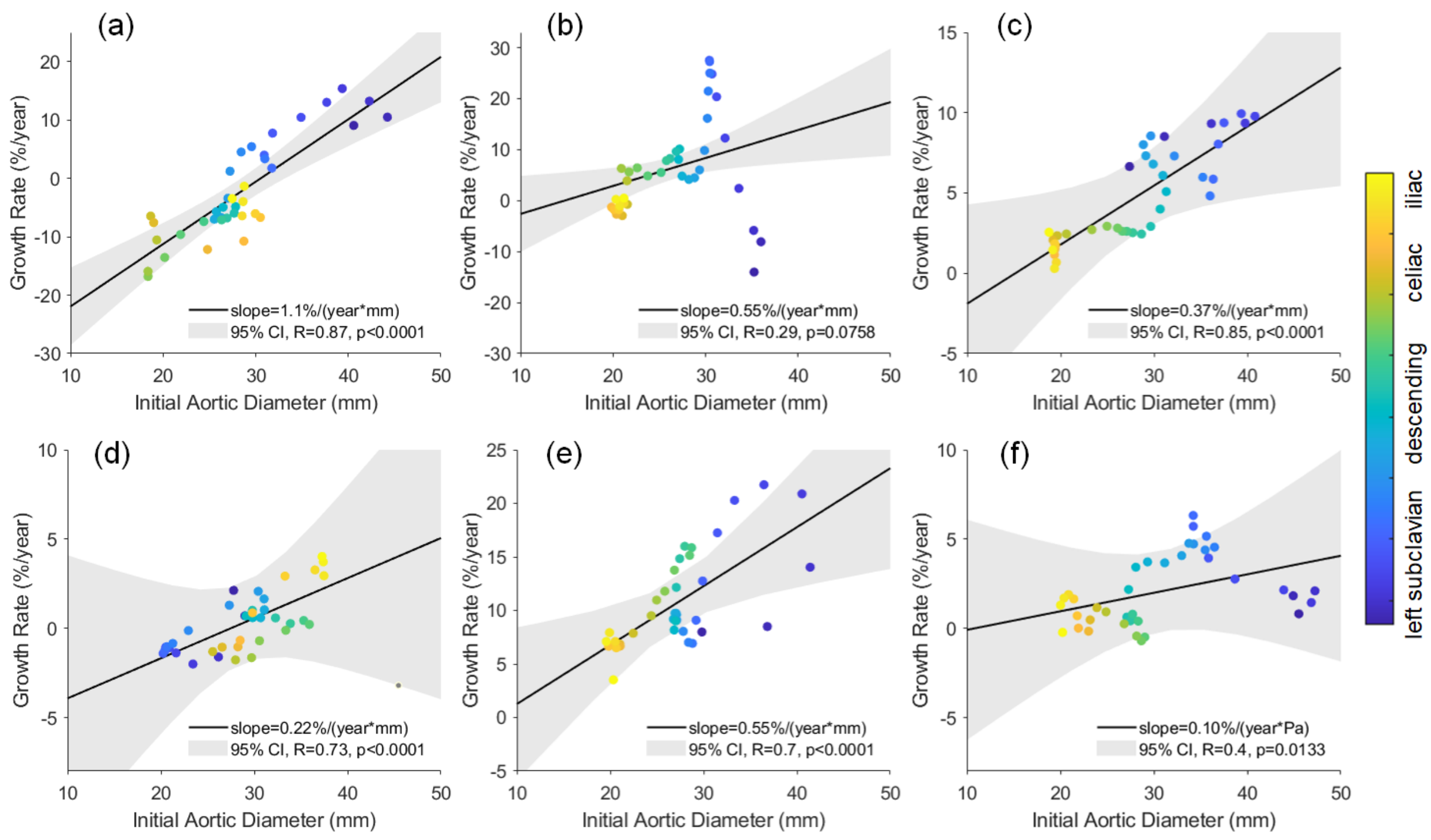

**[Figure A4. Initial aortic diameter of baseline geometry and aortic growth rate in 6 representative patients. (a) P12; (b) P14; (c) P15; (d) P17; (e) P22; (f) P25. Data points are color-coded by spatial location from the proximal to the distal aorta. The regression lines were obtained by using both fixed- and random-effects in the linear mixed-effects model. CI: confidence interval; R, p: Pearson correlation coefficient and its corresponding p-value.]**

*8.3. Classification analyses using WSS, pressure and initial diameter*

Figure A5 summarizes the individual correlations of classification analyses between WSS distributions and aortic diameter growth for representative patients (P12, P14, P15, P17, P22, and P25).

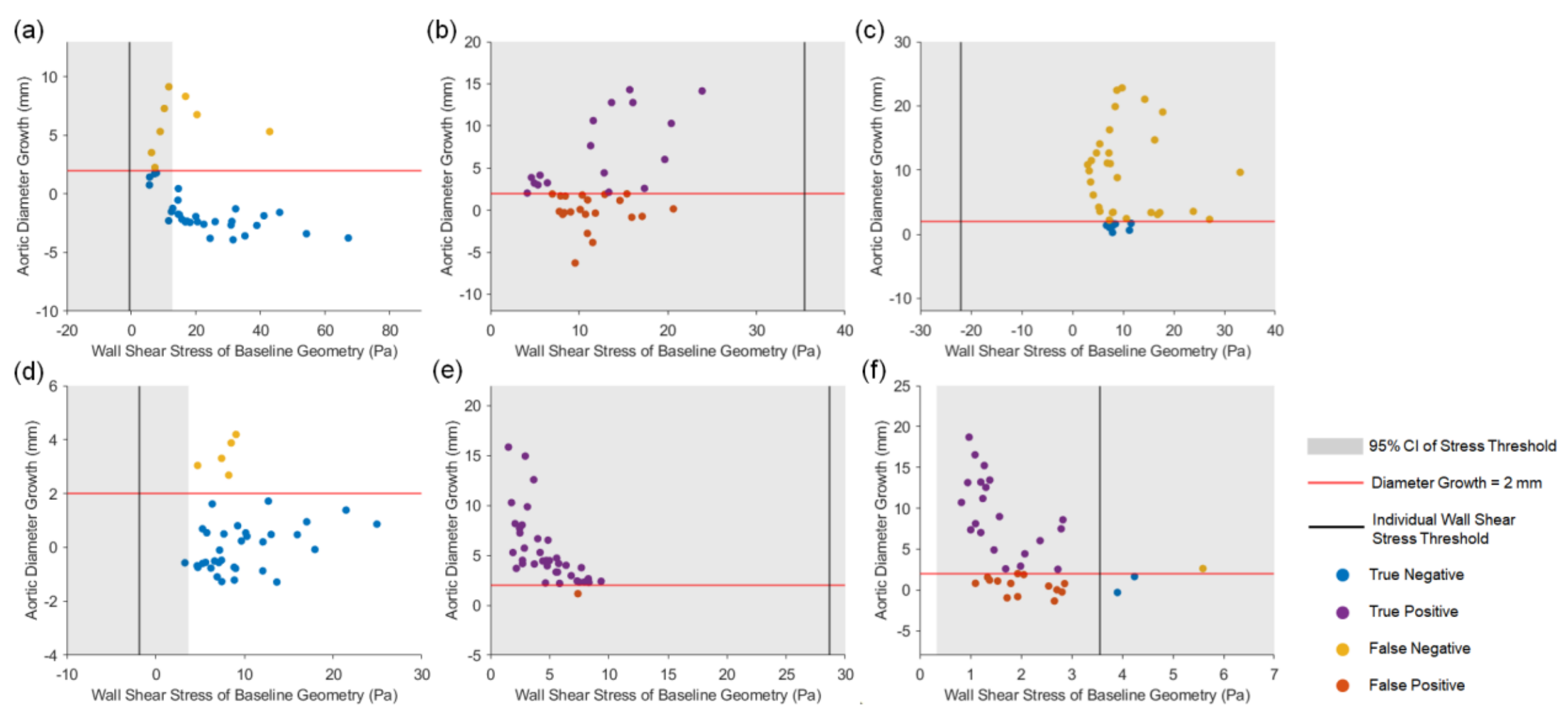

**[Figure A5. WSS of baseline geometry and aortic diameter growth in 6 representative patients and classification results using mixed-effects logistic regression. (a) P12, (b) P14, (c) P15, (d) P17, (e) P22, and (f) P25. Red: true negative, yellow: true positive, purple: false negative, and blue: false positive. Red horizontal line is the diameter change threshold = 2mm, and the black vertical line is the patient-specific WSS threshold obtained from the classification model. CI: confidence interval.]**

Figure A6 summarizes the individual correlations of classification analyses between pressure distributions and aortic diameter growth for representative patients (P12, P14, P15, P17, P22, and P25).

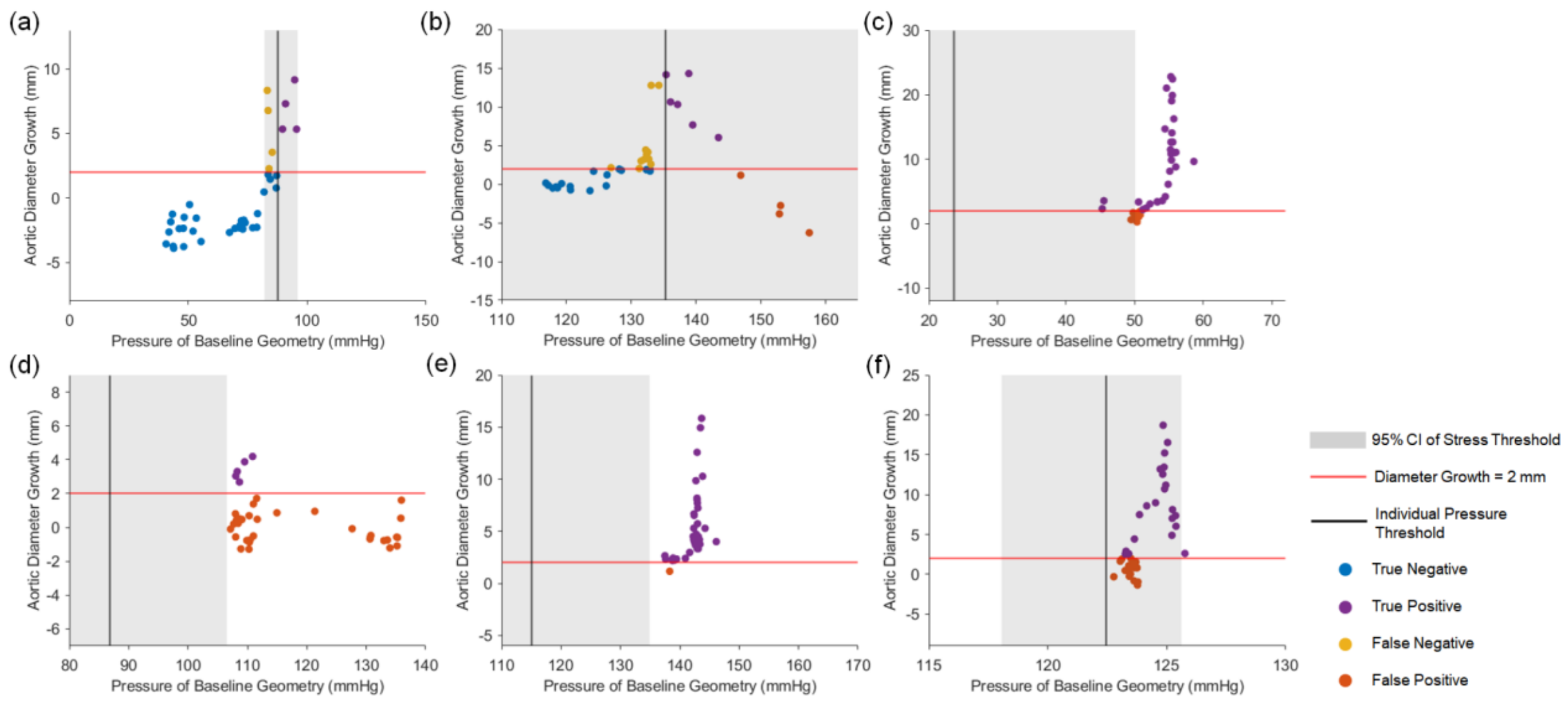

**[Figure A6. Pressure of baseline geometry and aortic diameter growth in 6 representative patients and classification results using mixed-effects logistic regression. (a) P12, (b) P14, (c) P15, (d) P17, (e) P22, and (f) P25. Red: true negative, yellow: true positive, purple: false negative, and blue: false positive. Red horizontal line is the diameter change threshold = 2mm, and the black vertical line is the patient-specific pressure threshold obtained from the classification model. CI: confidence interval.]**

Figure A7 summarizes the individual correlations of classification analyses between initial aortic diameter and aortic diameter growth for representative patients (P12, P14, P15, P17, P22, and P25).

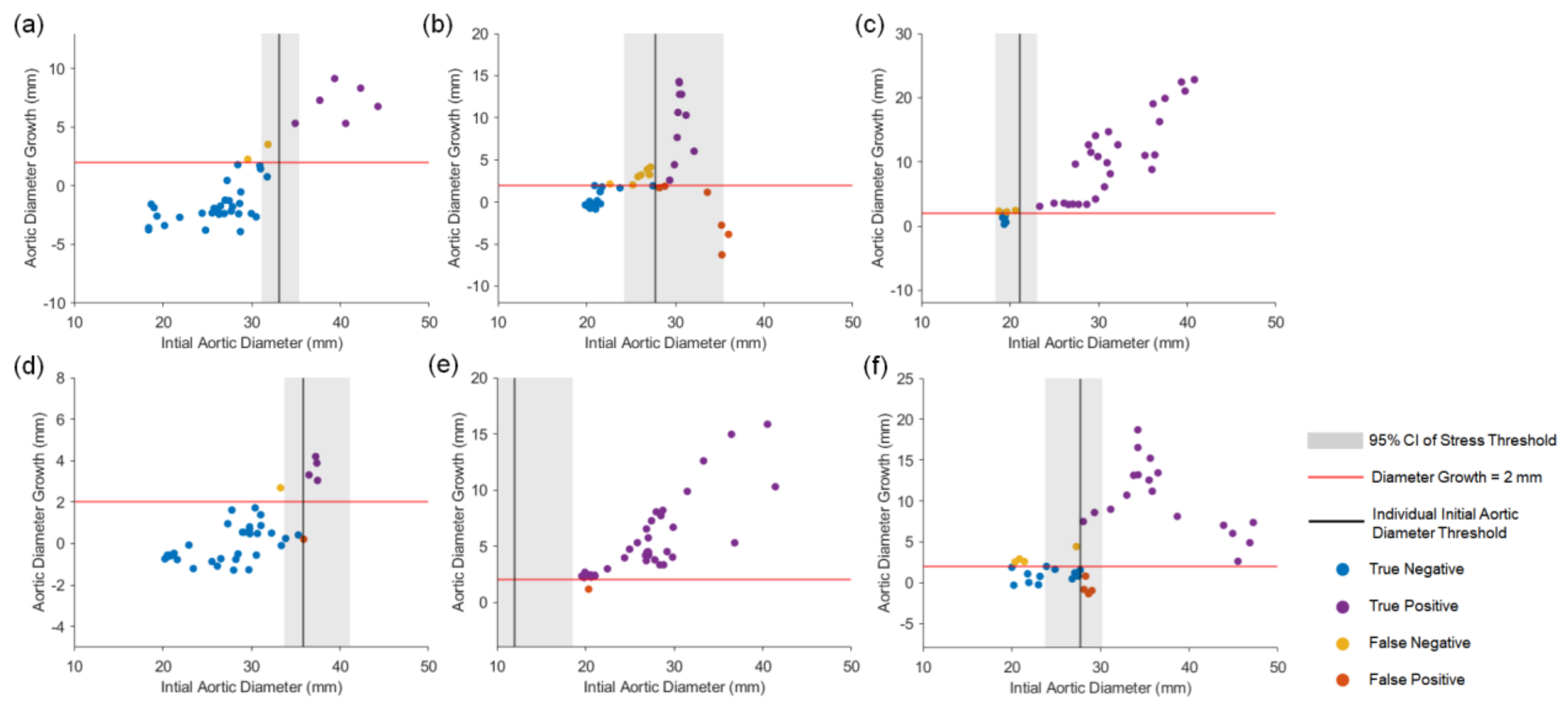

**[Figure A7. Initial aortic diameter and aortic diameter growth in 6 representative patients and classification results using mixed-effects logistic regression. (a) P12, (b) P14, (c) P15, (d) P17, (e) P22, and (f) P25. Red: true negative, yellow: true positive, purple: false negative, and blue: false positive. Red horizontal line is the diameter change threshold = 2mm, and the black vertical line is the patient-specific initial aortic diameter threshold obtained from the classification model. CI: confidence interval.]**